\documentclass[11pt]{article} % onecolumn (standard format)
\usepackage[left=2.5cm, right=2.5cm, top=1.785cm, bottom=2.0cm]{geometry}
\usepackage[utf8]{inputenc}
\usepackage{amsmath}
\usepackage{amsfonts}
\usepackage{amssymb}
\usepackage{mathtools}
\usepackage{listings}
\usepackage{mathrsfs}
\usepackage{times}
\usepackage{float}
\usepackage{color}
\usepackage{setspace}
\usepackage{color,soul}

\usepackage{amsmath,amsfonts,amsthm,eucal}
\usepackage{enumerate}
\usepackage{color}
\usepackage{tabularx}
\usepackage{siunitx}
\usepackage{empheq}
\usepackage[tableposition=below]{caption}
\usepackage[
pdfstartview=XYZ,
bookmarks=true,
colorlinks=true,
linkcolor=blue,
urlcolor=blue,
citecolor=blue,
pdftex,
bookmarks=true,
linktocpage=true,   % makes the page number as hyperlink in table of content
hyperindex=true
]{hyperref}

\usepackage{lipsum}
\usepackage[FIGBOTCAP,TABTOPCAP,tight]{subfigure}
\usepackage{subcaption}
 
\usepackage[pdftex]{graphicx}
\usepackage{epstopdf}
\usepackage{booktabs} 
\usepackage{tikz,mathpazo}
\usetikzlibrary{shapes.geometric, arrows}
\usepackage{caption}

\usepackage{longtable}
\usepackage{adjustbox}
\usepackage{framed}

\usepackage[colorinlistoftodos]{todonotes}

\usepackage{amsmath}

\usepackage{nicefrac}
\usepackage{multirow}
\usepackage{hhline}
\usepackage{lineno}
\usepackage{import}
\usepackage{epstopdf}
\usepackage{subcaption}
\usepackage{mathtools}
\usepackage{transparent}

\usepackage{cancel}
\usepackage{empheq}
\usepackage{booktabs}
\usepackage{cleveref}

\crefname{equation}{Eq.}{Eqs.}
\usepackage{soul}
\usepackage{courier}

\definecolor{tbf}{RGB}{255,0,0} % to be fixed
\definecolor{txue}{RGB}{0,0,255}

\usepackage{authblk}
\usepackage{listings}
\usepackage{titlesec}
\usepackage[numbers,sort&compress]{natbib}

\usepackage[
pdfstartview=XYZ,
bookmarks=true,
colorlinks=true,
linkcolor=blue,
urlcolor=blue,
citecolor=blue,
pdftex,
bookmarks=true,
linktocpage=true,
hyperindex=true
]{hyperref}

\newcommand\keywords[1]{\textbf{Keywords}: #1}

\title{\Large Style-constrained inverse design of microstructures with tailored mechanical properties using unconditional diffusion models} 

\begin{document}

\author[1]{\normalsize Weipeng Xu}
\author[1]{\normalsize Ziyuan Xie}
\author[1] {\normalsize Haoju Lin}
\author[1]{\normalsize Xinyu Wang}
\author[1]{\normalsize Guangjin Mou}
\author[1]{\normalsize Tianju Xue
\footnote{\textit{cetxue@ust.hk}  (corresponding author)}}
\affil[1]{\footnotesize Department of Civil and Environmental Engineering, Hong Kong University of Science and Technology, Hong Kong, China}

\date{}
\maketitle

\vspace{-30pt}

\begin{abstract}
Deep generative models, particularly denoising diffusion models, have achieved remarkable success in high-fidelity generation of architected microstructures with desired properties and styles. 
Nevertheless, these recent methods typically rely on conditional training mechanisms and demand substantial computational effort to prepare the labeled training dataset, which makes them inflexible since any change in the governing equations or boundary conditions requires a complete retraining process. 
In this study, we propose a new inverse design framework that integrates unconditional denoising diffusion models with differentiable programming techniques for architected microstructure generation.
Our approach eliminates the need for expensive labeled dataset preparation and retraining for different problem settings. 
By reinterpreting the noise input to the diffusion model as an optimizable design variable, we formulate the design task as an optimization problem over the noise input, enabling control over the reverse denoising trajectory to guide the generated microstructure toward the desired mechanical properties while preserving the stylistic constraints encoded in the training dataset. 
A unified differentiation pipeline via vector-Jacobian product concatenations is developed to enable end-to-end gradient evaluation through backpropagation. 
Several numerical examples, ranging from the design of microstructures with specified homogenized properties to those with targeted hyperelastic and elasto-plastic behaviors, showcase the effectiveness of the framework and its potential for advanced design tasks involving diverse performance and style requirements.
\end{abstract}


\section{Introduction}
\label{sec:intro}
Architected microstructures have gained wide adoption in advanced engineering scenarios~\cite{kazim2025mechanical,zheng2025metamaterial} due to their superior and tunable mechanical properties. 
Their intricate geometric configurations allow them to break the inherent limitations of the base material, enabling a range of novel functionalities, such as negative Poisson's ratios~\cite{zhang2025mechanical}, energy absorption~\cite{zeng2023inverse}, and strain cloaking~\cite{wang2022mechanical}. The design task of architected microstructures involves two distinct yet interrelated aspects: performance and style. 
Performance refers to the desired mechanical behaviors or properties that the microstructure should exhibit, such as specific stiffness, strength, or deformation characteristics. 
Style, on the other hand, pertains to the geometric features and aesthetic qualities of the microstructure, which may be influenced by geometric and manufacturing constraints~\cite{maurizi2025designing}, or visual appeal~\cite{sampaio2022lightweight}. 
Achieving a harmonious balance between performance and style is crucial for the successful implementation of architected microstructures in practical applications. 

As for the design methodologies, while heuristic-based direct design~\cite{overvelde2012compaction,shan2015multistable} and optimization-based inverse design~\cite{zeng2023inverse,sha2024topology} have been widely explored to tailor the performance of microstructures, they often require additional efforts to incorporate stylistic constraints~\cite{navez2022topology,zhang2023machine}.
Recently, data-driven approaches~\cite{lee2024data,flaschel2023automated} have established a new paradigm for the computational analysis and design of advanced materials and structures. 
Among these, deep generative models, such as variational autoencoders (VAE)~\cite{kingma2013auto}, generative adversarial networks (GAN)~\cite{goodfellow2014generative}, flow-based models~\cite{dinh2016density}, and denoising diffusion models~\cite{ho2020denoising,song2020denoising,song2020score}, are emerging as powerful tools for the style-constrained inverse design of microstructures with tailored mechanical properties. 
Early works~\cite{wang2020deep,li2024design,mao2020designing,zheng2021controllable,wang2022ih} primarily focused on the use of VAE and GAN for microstructure design. 
More recently, denoising diffusion models and flow-based models have been investigated for this task due to their superior generation quality and diversity. 

Recent works typically employ \textit{conditional} diffusion models to steer the generation process toward the desired properties while preserving the stylistic features encoded in the training dataset. 
For instance, Vlassis and Sun~\cite{vlassis2023denoising} developed a conditional denoising diffusion algorithm to generate hyperelastic microstructures with specified stress-strain responses and handwritten digit styles, where the generation process is guided by embedded feature vectors that encode target mechanical behaviors. 
Bastek and Kochmann~\cite{bastek2023inverse} proposed a video diffusion generative models conditioned on target stress-strain curves for the inverse design of periodic stochastic cellular structures with tuned nonlinear deformation and stress response under large-strain compression. 
Wang et al.~\cite{wang2024diffmat} presented a conditional denoising diffusion framework \texttt{DiffMat} for the design of energy-absorbing microstructures, where the target stress-strain response is encoded via a cross-attention mechanism. 
Li et al.~\cite{LI2025121417} leveraged a conditional denoising diffusion model to generate composite microstructures with target hyperelastic properties. The strain energy value and reaction forces are employed as the condition inputs for the training and generation processes. 
Zhang et al.~\cite{zhang2025hyperdiff} developed a conditional diffusion model \texttt{HyperDiff} for the inverse design of hyperelastic porous microstructures with targeted force-displacement curves. 
The simulated force-displacement curves are smoothed and parameterized using B-spline curves, and the normalized coefficients and knots are employed as the conditional inputs for the diffusion model. 

Besides, some recent works have explored the use of advanced model architectures and extra guidance techniques to enhance the generation quality and efficiency. Mirzaee et al.~\cite{mirzaee2025inverse} introduced a flow-based conditional generative design framework \texttt{PoreFlow} for porous material microstructures with specified physical properties. 
Yang et al.~\cite{yang2025graphdgm} proposed \texttt{GraphDGM}, a graph-based conditional  diffusion model for the design of frame and lattice structures. Zhang et al.~\cite{zhang2025conditional} introduced a conditional diffusion model \texttt{LatticeOptDiff} for the design of lattice structures with tailored elastic properties and volume fractions, where the generation process is guided by a property regression network to enhance the design accuracy. 
Zheng et al.~\cite{zheng2025diffumeta} proposed a conditional inverse design framework \texttt{DiffuMeta} for shell-type structures with target elastic properties and stress-strain responses under large deformations. An algebraic language-based parameterization is introduced to encode complex geometries as mathematical token sequences and integrated into a diffusion transformer architecture to enable high-quality generation. 

These aforementioned conditional generative design methods must rely on extensive datasets with labeled property information for training. 
The labeled data are typically obtained through iterative numerical simulations.
% While some of them can leverage existing mechanical datasets, most of them require computationally expensive simulations to generate labeled data for each specific problem setting. 
Any change in the governing equations or boundary conditions necessitates a complete regeneration of the labeled dataset and retraining of the generative model, which limits their flexibility and applicability for diverse design scenarios.

In this study, we propose a new inverse design framework that integrates \textit{unconditional} denoising diffusion models with advanced differentiable programming techniques~\cite{blondel2024elements} for architected microstructures with tailored mechanical properties and stylistic constraints.
Our approach eliminates the need for expensive labeled dataset preparation and model retraining for different problem settings.
The key idea of our approach is to reinterpret the noise input to the diffusion model as an optimizable design variable that encodes the geometric features of the target microstructure. 
By formulating the design task as an optimization problem over the noise input, we can control the reverse denoising trajectory to guide the generated microstructure toward the desired mechanical properties. 
The stylistic constraints encoded in the training dataset are inherently preserved during the optimization process, ensuring that the resulting microstructures maintain the desired geometric features. 
The algorithms are implemented through a unified differentiation pipeline via vector-Jacobian product (VJP) concatenations to enable end-to-end gradient evaluation through backpropagation~\cite{rumelhart1986learning,griewank2008evaluating}, allowing the use of gradient-based optimization algorithms to efficiently solve the inverse design problem.

The rest of this paper is organized as follows. In Section~\ref{sec:method}, we introduce the proposed style-constrained inverse design framework for architected microstructures with tailored mechanical properties. In Section~\ref{sec:examples}, we demonstrate the effectiveness of the proposed method through several numerical examples in different mechanical scenarios, including the design of microstructures with specified homogenized, hyperelastic, and elasto-plastic properties. A discussion on the advantages, limitations, and potential future directions of the proposed method is also included. Finally, we conclude the paper in Section~\ref{sec:conclusion}.

\section{Inverse design framework using unconditional diffusion models}
\label{sec:method}
In this section, we begin by providing a brief overview of denoising diffusion probabilistic models (DDPM)~\cite{ho2020denoising} and denoising diffusion implicit models (DDIM)~\cite{song2020denoising}. Next, the details of the proposed inverse design algorithm are explained, followed by the gradient evaluation procedure via a unified differentiation pipeline using vector-Jacobian product concatenations. Finally, a conceptual example of Gaussian mixture models is presented to illustrate the core mechanism of our approach.

\subsection{Denoising diffusion models}
\label{method:diffusion}
The standard workflow of DDPM consists of two processes: a forward diffusion process that gradually adds noise to the data and a reverse denoising process that learns to remove the noise and recover the original data distribution. 

In the forward diffusion process, given a sample $\boldsymbol{x}_0$ drawn from the data distribution $q(\boldsymbol{x}_0)$, a series of latents $\boldsymbol{x}_{t}$ for $t=1, \ldots, T$ are generated by progressively adding Gaussian noise over $T$ time steps. This process can be defined as a Markov chain
\begin{align}
  q(\boldsymbol{x}_{1:T} \mid \boldsymbol{x}_0) = \prod_{t=1}^T q(\boldsymbol{x}_t \mid \boldsymbol{x}_{t-1}),
\end{align}
where the transition kernel $q(\boldsymbol{x}_t \mid \boldsymbol{x}_{t-1})$ is given by
\begin{align}
  q(\boldsymbol{x}_t \mid \boldsymbol{x}_{t-1}) = \mathcal{N}(\boldsymbol{x}_t; \sqrt{1 - \beta_t} \boldsymbol{x}_{t-1}, \beta_t \mathbf{I}), 
\end{align}
with $\beta_t\in(0,1)$ being a variance schedule that controls the amount of noise added at each time step. This process gradually transforms the data distribution into an isotropic Gaussian distribution $\mathcal{N}(\mathbf{0}, \mathbf{I})$ as $t$ approaches $T$. The forward process can be further expressed in a closed form as
\begin{align}
  q(\boldsymbol{x}_t \mid \boldsymbol{x}_0) = \mathcal{N}(\boldsymbol{x}_t; \sqrt{\bar{\alpha}_t} \boldsymbol{x}_0, (1 - \bar{\alpha}_t) \mathbf{I}),
\end{align}
where $\alpha_t = 1 - \beta_t$ and $\bar{\alpha}_t = \prod_{s=1}^t \alpha_s$. Then, we can directly sample $\boldsymbol{x}_t$ at any time step $t$ from $\boldsymbol{x}_0$ and a sampled noise $\boldsymbol{\epsilon} \sim \mathcal{N}(\mathbf{0}, \mathbf{I})$ as
\begin{align}
  \boldsymbol{x}_t = \sqrt{\bar{\alpha}_t} \boldsymbol{x}_0 + \sqrt{1 - \bar{\alpha}_t} \boldsymbol{\epsilon}.
\end{align}

The reverse denoising process aims to learn a parameterized transition $p_\theta(\boldsymbol{x}_{t-1} \mid \boldsymbol{x}_t)$ that gradually removes the noise from $\boldsymbol{x}_T$ to recover the original data $\boldsymbol{x}_0$. This process can be defined as
\begin{align}
  \label{eq:diff_rev}
  p_\theta(\boldsymbol{x}_{0:T}) = p(\boldsymbol{x}_T) \prod_{t=1}^T p_\theta(\boldsymbol{x}_{t-1} \mid \boldsymbol{x}_t),
\end{align}
where $p(\boldsymbol{x}_T) = \mathcal{N}(\boldsymbol{x}_T; \mathbf{0}, \mathbf{I})$ is the prior distribution, serving as the starting point for the reverse process. The transition kernel $p_\theta(\boldsymbol{x}_{t-1} \mid \boldsymbol{x}_t)$ is parameterized as
\begin{align}
  p_\theta(\boldsymbol{x}_{t-1} \mid \boldsymbol{x}_t) = \mathcal{N}(\boldsymbol{x}_{t-1}; \boldsymbol{\mu}_\theta(\boldsymbol{x}_t, t), \boldsymbol{\Sigma}_{\theta}(\boldsymbol{x}_t, t)),
\end{align}
with the mean $\boldsymbol{\mu}_\theta(\boldsymbol{x}_t, t)$ as
\begin{align}
  \boldsymbol{\mu}_\theta(\boldsymbol{x}_t, t) = \frac{1}{\sqrt{\alpha_t}} \left( \boldsymbol{x}_t - \frac{\beta_t}{\sqrt{1 - \bar{\alpha}_t}} \boldsymbol{\epsilon}_\theta(\boldsymbol{x}_t, t) \right),
\end{align}
where $\boldsymbol{\epsilon}_\theta$ is a denoising neural network with parameters $\theta$ that predicts the added noise at each time step.
The variance $\boldsymbol{\Sigma}_\theta$ can be modeled as a fixed or learnable parameter. We follow the original DDPM paper~\cite{ho2020denoising} and set it as $\boldsymbol{\Sigma}_\theta(\boldsymbol{x}_t, t) = \tilde{\beta}_t \mathbf{I}$, where $\tilde{\beta}_t$ is defined as
\begin{align}
  \tilde{\beta}_t = \frac{1 - \bar{\alpha}_{t-1}}{1 - \bar{\alpha}_t} \beta_t.
\end{align}
The denoising neural network is trained by minimizing the following objective function:
\begin{align}
  \min_\theta \mathbb{E}_{t, \boldsymbol{x}_0, \boldsymbol{\epsilon}} \left[ \left\| \boldsymbol{\epsilon} - \boldsymbol{\epsilon}_\theta(\sqrt{\bar{\alpha}_t} \boldsymbol{x}_0 + \sqrt{1 - \bar{\alpha}_t} \boldsymbol{\epsilon}, t) \right\|^2_2 \right].
\end{align} 
The trained model can generate new samples by starting from a random noise $\boldsymbol{x}_T\sim p(\boldsymbol{x}_T)$ and iteratively applying the learned reverse transition $p_\theta(\boldsymbol{x}_{t-1} \mid \boldsymbol{x}_t)$ to obtain $\boldsymbol{x}_0$. For each time step $t$, the sampling process can be expressed as
\begin{align}
  \label{eq:ddpm_rev}
\boldsymbol{x}_{t-1} = \frac{1}{\sqrt{\alpha_t}} \left( \boldsymbol{x}_t - \frac{1-\alpha_t}{\sqrt{1 - \bar{\alpha}_t}} \boldsymbol{\epsilon}_\theta(\boldsymbol{x}_t, t) \right) + \sigma_t \boldsymbol{z},
\end{align}
where $\sigma_t = \sqrt{\tilde{\beta}_t}$ and $\boldsymbol{z} \sim \mathcal{N}(\mathbf{0}, \mathbf{I})$ is the random noise added at each step. However, sampling with the Markovian process~(\ref{eq:ddpm_rev}) requires a large number of time steps to achieve high-quality results, which can be computationally expensive. 

To address this issue, DDIM~\cite{song2020denoising} proposes a non-Markovian reverse process that enables faster sampling while maintaining sample quality. The forward process of DDIM remains the same as DDPM, while the reverse process is modified to
\begin{align}
  \label{eq:ddim}
  \boldsymbol{x}_{\tau_{i-1}} = \sqrt{\bar{\alpha}_{\tau_{i-1}}} \left( \frac{\boldsymbol{x}_{\tau_i} - \sqrt{1 - \bar{\alpha}_{\tau_i}} \boldsymbol{\epsilon}_\theta(\boldsymbol{x}_{\tau_i}, \tau_i)}{\sqrt{\bar{\alpha}_{\tau_i}}} \right) + \sqrt{1 - \bar{\alpha}_{\tau_{i-1}} - \sigma_{\tau_i}(\eta)^2} \boldsymbol{\epsilon}_\theta(\boldsymbol{x}_{\tau_i}, \tau_i) + \sigma_{\tau_i}(\eta) \boldsymbol{z},
\end{align}
where the time steps $\{\tau_i\}$ form a subsequence $S\subseteq\{1, \ldots, T\}$ and can be reduced to a smaller number compared to $T$ for faster sampling.
$\sigma_{\tau_i}(\eta)$ is a noise scale controlled by the hyperparameter $\eta\in[0,1]$. 
By setting $\eta=0$ (i.e., $\sigma_{\tau_i}=0$), the sampling process follows a fixed trajectory and becomes deterministic. 
This one-to-one mapping between the initial noise and the generated sample, together with the sampling efficiency, makes DDIM particularly suitable for traceable design tasks. 
Therefore, we adopt DDIM with $\eta=0$ as the generative model in this work.

\subsection{Inverse design method}
\label{method:algorithm}
Existing inverse design approaches based on diffusion models typically condition the reverse denoising process~(\ref{eq:diff_rev}) on target mechanical properties.
In this study, we propose an inverse design framework that integrates unconditional denoising diffusion models with differentiable programming techniques to achieve style-constrained design of microstructures with tailored mechanical properties.
Our framework eliminates the need for expensive labeled dataset preparation and model retraining if a new problem setting is considered.

\begin{figure}[H] \centering
    {\includegraphics[width=1\textwidth]{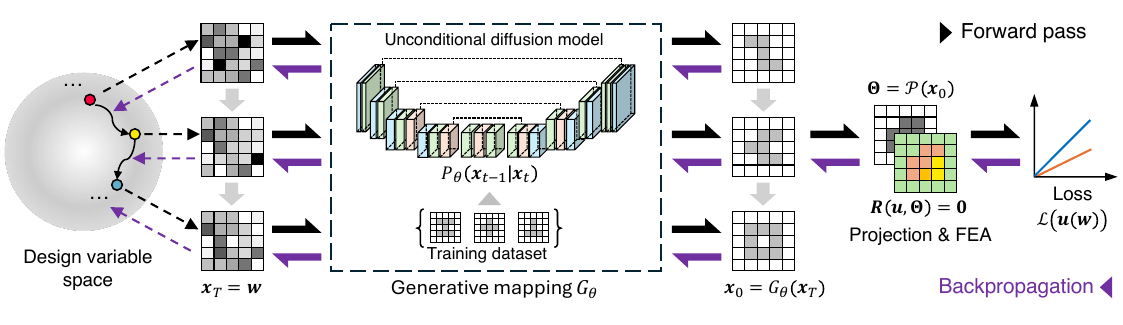}}
    \caption{Schematic illustration of the style-constrained inverse design framework for microstructures with tailored mechanical properties using unconditional diffusion models.
    } \label{fig:framework}
\end{figure}

Fig.~\ref{fig:framework} illustrates the workflow of the proposed inverse design framework, which consists of two main steps: training an unconditional diffusion model on an image dataset and optimizing the noise input to generate target microstructures via backpropagation.

The first step is to train an unconditional denoising diffusion model with  neural network parameters $\theta$ on the prepared image dataset, following the procedure outlined in Section~\ref{method:diffusion}. 
The trained diffusion model captures the geometric characteristics and stylistic 
constraints encoded in the training dataset and enables the generation 
of new samples that inherit these learned features.
Following the DDIM formulation in Eq.~(\ref{eq:ddim}) with $\eta=0$, we obtain a deterministic mapping $\mathcal{G}_\theta: \mathbb{R}^{n} \rightarrow \mathbb{R}^{n}$ that maps the noise input $\boldsymbol{x}_T\in\mathbb{R}^{n}$ 
to a microstructure image $\boldsymbol{x}_0\in\mathbb{R}^{n}$ through the learned denoising process, where $n$ denotes the image dimension.

Next, we formulate the microstructure design task as a constrained optimization problem with the noise input as the design variable $\boldsymbol{w}\in \mathbb{R}^{n}$. In other words, we seek to find an optimal noise input $\boldsymbol{w}^*$ such that the generated microstructure exhibits the desired mechanical properties. To evaluate the mechanical response of the generated microstructure, we first convert the image $\boldsymbol{x}_0= \mathcal{G}_\theta(\boldsymbol{w})$ into a finite element mesh and assign material properties to each element based on the pixel intensity values using a projection function. Then, we perform finite element analysis (FEA) to compute the mechanical response of the microstructure under specified boundary conditions. Next, the loss function $\mathcal{L}$ is evaluated to quantify the discrepancy between the current and target mechanical properties. 
The design objective is to minimize $\mathcal{L}$ subject to the equilibrium equations of the mechanical system as well as the material property field sourced from the diffusion model-based generation. This leads to the following optimization problem:
\begin{align}
\label{eq:opt}
      &\min_{\ \boldsymbol{w}\in\mathbb{R}^{n}}\mathcal{L}\left(\boldsymbol{u}\left(\boldsymbol{w}\right)\right)\nonumber\\
    &\textrm{s.t.}\quad \boldsymbol{R}\left(\boldsymbol{u},\boldsymbol{\Theta}\right)=\boldsymbol{0},\\
    &\quad\quad \boldsymbol{\Theta}=\mathcal{P}\left(\boldsymbol{x}_0\right),\nonumber\\
    &\quad\quad\boldsymbol{x}_0=\mathcal{G}_\theta(\boldsymbol{w}),\nonumber
\end{align}
where $\boldsymbol{u}\in\mathbb{R}^{m}$ is the mechanical response of the microstructure, $\boldsymbol{R}: \mathbb{R}^{m}\times\mathbb{R}^{n}\rightarrow\mathbb{R}^{m}$ represents the residual function obtained from  discretizing the governing equations of the mechanical system, $\boldsymbol{\Theta}\in\mathbb{R}^{n}$ denotes the material property field assigned to the microstructure, and $\mathcal{P}: \mathbb{R}^{n}\rightarrow\mathbb{R}^{n}$ is a projection function that maps the generated image $\boldsymbol{x}_0=\mathcal{G}_\theta(\boldsymbol{w})$ to $\boldsymbol{\Theta}$.
The cornerstone of this approach lies in the reinterpretation of the noise input: rather than treating it as a disposable initial state for each generation, we reconceptualize it as an optimizable design variable that encodes the geometric features of the target microstructure. This reformulation transforms the typical generative workflow into an optimization-driven inverse design process. 

To solve this optimization problem, the gradient information, i.e., $\frac{\mathrm{d}{\mathcal{L}}}{\mathrm{d}\boldsymbol{w}}$ is crucial.
We establish a unified differentiation pipeline via vector-Jacobian product concatenations, as detailed in Section~\ref{method:grad}, to allow the mechanical response $\boldsymbol{u}$ to be differentiable with respect to the material property field $\boldsymbol{\Theta}$. 
This end-to-end differentiability of the entire framework, comprising the deterministic diffusion model and FEA, enables gradient evaluations through automatic differentiation~\cite{griewank2008evaluating}. Specifically, the gradient $\frac{\mathrm{d}{\mathcal{L}}}{\mathrm{d}\boldsymbol{w}}$ is efficiently obtained via backpropagation~\cite{rumelhart1986learning} through the entire framework, allowing us to employ gradient-based optimization algorithms to solve the optimization problem~(\ref{eq:opt}).

\subsection{Unified differentiation pipeline via vector-Jacobian product concatenations}
\label{method:grad}
In this section, we present the details for evaluating the gradient of the loss function $\mathcal{L}$ with respect to the design variable $\boldsymbol{w}$, i.e., $\frac{\mathrm{d}{\mathcal{L}}}{\mathrm{d}\boldsymbol{w}}$, through a unified differentiation pipeline. In the context of automatic differentiation~\cite{griewank2008evaluating}, we can decompose the forward computation process into a sequence of elementary operations, each represented as a function with well-defined inputs and outputs. The resulting computational graph captures the dependencies among these operations, enabling systematic gradient evaluation via the chain rule. Following this principle, we can identify three key components in our framework: the genreative mapping $\mathcal{G}_\theta$, the projection function $\mathcal{P}$, and the mechanical solver for $\boldsymbol{\boldsymbol{u}}$. By applying the chain rule, the total gradient of the loss function $\mathcal{L}$ with respect to the design variable $\boldsymbol{w}$ can be expressed as
\begin{align}
  \label{eq:chain}
  \frac{\mathrm{d} \mathcal{L}}{\mathrm{d} \boldsymbol{w}} = \frac{\partial \mathcal{L}}{\partial \boldsymbol{u}} \frac{\partial \boldsymbol{u}}{\partial \boldsymbol{\Theta}} \frac{\partial \boldsymbol{\Theta}}{\partial \boldsymbol{x}_0} \frac{\partial \boldsymbol{x}_0}{\partial \boldsymbol{w}},
\end{align}
where each term is not computed explicitly as a full Jacobian matrix, but rather through vector-Jacobian products (VJP) that efficiently propagate gradients backward through the computational graph. Let $\boldsymbol{v}_{\left(*\right)}=\left[\frac{\partial \mathcal{L}}{\partial \left(*\right)}\right]^\top$ denotes the propagated vector at the intermediate variable $\left(*\right)$. Then, the backpropagation can be performed by sequentially computing the VJPs as follows:
\begin{subequations}
\begin{align}
  \boldsymbol{v}_{\boldsymbol{u}}^\top &= \frac{\partial \mathcal{L}}{\partial \boldsymbol{u}}\label{eq:vjp_u}, \\
  \boldsymbol{v}_{\boldsymbol{\Theta}}^\top &= \boldsymbol{v}_{\boldsymbol{u}}^\top \frac{\partial \boldsymbol{u}}{\partial \boldsymbol{\Theta}}\label{eq:vjp_theta}, \\
  \boldsymbol{v}_{\boldsymbol{x}_0}^\top &= \boldsymbol{v}_{\boldsymbol{\Theta}}^\top \frac{\partial \boldsymbol{\Theta}}{\partial \boldsymbol{x}_0}\label{eq:vjp_g},
\end{align}
\end{subequations}
and the total gradient $\frac{\mathrm{d} \mathcal{L}}{\mathrm{d} \boldsymbol{w}}$ can be obtained as follows: 
\begin{align}
  \label{eq:vjp_w}
  \frac{\mathrm{d} \mathcal{L}}{\mathrm{d} \boldsymbol{w}} = \boldsymbol{v}_{\boldsymbol{x}_0}^\top \frac{\partial \boldsymbol{x}_0}{\partial \boldsymbol{w}}.
\end{align}
Among these VJP components, (\ref{eq:vjp_u}), (\ref{eq:vjp_g}), and (\ref{eq:vjp_w}) can be readily computed using standard backpropagation algorithms due to the differentiability of the neural network, projection function, and loss function. The critical part lies in evaluating the $\boldsymbol{v}_{\boldsymbol{u}}^\top \frac{\partial \boldsymbol{u}}{\partial \boldsymbol{\Theta}}$ in Eq.~(\ref{eq:vjp_theta}), which involves the sensitivity of the mechanical response $\boldsymbol{u}$ with respect to the material property field $\boldsymbol{\Theta}$. This is not straightforward since $\boldsymbol{u}$ is implicitly defined by the equilibrium equations of the mechanical system. 

To find  $\boldsymbol{v}_{\boldsymbol{u}}^\top \frac{\partial \boldsymbol{u}}{\partial \boldsymbol{\Theta}}$, we employ the implicit differentiation framework~\cite{blondel2022efficient} to derive adjoint-based VJP rules for the mechanical response $\boldsymbol{u}$, enabling efficient gradient propagation through the mechanical solver via the adjoint analysis~\cite{givoli2021tutorial,cao2003adjoint}. We start by differentiating the equilibrium equation $\boldsymbol{R}(\boldsymbol{u}, \boldsymbol{\Theta}) = \mathbf{0}$ with respect to $\boldsymbol{\Theta}$
\begin{align}
  \frac{\partial \boldsymbol{R}}{\partial \boldsymbol{u}} \frac{\partial \boldsymbol{u}}{\partial \boldsymbol{\Theta}} + \frac{\partial \boldsymbol{R}}{\partial \boldsymbol{\Theta}} = \mathbf{0}.
\end{align}
Rearranging the above equation yields
\begin{align}
  \label{eq:implicit}
  \frac{\partial \boldsymbol{u}}{\partial \boldsymbol{\Theta}} = - \left( \frac{\partial \boldsymbol{R}}{\partial \boldsymbol{u}} \right)^{-1} \frac{\partial \boldsymbol{R}}{\partial \boldsymbol{\Theta}},
\end{align}
which means the implicit derivative $\frac{\partial \boldsymbol{u}}{\partial \boldsymbol{\Theta}}$ can be computed via multiplications of explicit derivatives on the right-hand side. Then we multiply $\boldsymbol{v}_{\boldsymbol{u}}^\top$ to both sides of Eq.~(\ref{eq:implicit}) to obtain
\begin{align}
\label{eq:vjp_intermediate}
\boldsymbol{v}_{\boldsymbol{u}}^\top \frac{\partial \boldsymbol{u}}{\partial \boldsymbol{\Theta}} = - \boldsymbol{v}_{\boldsymbol{u}}^\top \left( \frac{\partial \boldsymbol{R}}{\partial \boldsymbol{u}} \right)^{-1} \frac{\partial \boldsymbol{R}}{\partial \boldsymbol{\Theta}}.
\end{align}
To avoid computing the inverse of $\frac{\partial \boldsymbol{R}}{\partial \boldsymbol{u}}$ explicitly, we introduce an adjoint variable $\boldsymbol{\lambda}_{\boldsymbol{u}}\in\mathbb{R}^{m}$ obtained by solving the following adjoint equation:
\begin{align}
\label{eq:adjoint}
\left( \frac{\partial \boldsymbol{R}}{\partial \boldsymbol{u}} \right)^\top \boldsymbol{\lambda}_{\boldsymbol{u}} = - \boldsymbol{v}_{\boldsymbol{u}}.
\end{align}
Substituting $\boldsymbol{\lambda}_{\boldsymbol{u}}$ back into Eq.~(\ref{eq:vjp_intermediate}) allows us to define the customized VJP rule for the mechanical solver as
\begin{align}
\label{eq:vjp}
\boldsymbol{v}_{\boldsymbol{u}}^{\top}\frac{\partial\boldsymbol{u}}{\partial\boldsymbol{\Theta}} &= \boldsymbol{\lambda}_{\boldsymbol{u}}^{\top}\frac{\partial\boldsymbol{R}}{\partial\boldsymbol{\Theta}},
\end{align}
which only involves the computation of the explicit derivative $\frac{\partial \boldsymbol{R}}{\partial \boldsymbol{\Theta}}$ and the solution of a linear system~(\ref{eq:adjoint}). The customized VJP rule (\ref{eq:vjp}) does not rely on the specific form of the residual function $\boldsymbol{R}$, making it applicable to a wide range of mechanical systems. By concatenating this customized VJP rule with the standard differentiation rules for the other VJP components, we establish a unified differentiation pipeline that enables efficient gradient evaluation through the entire inverse design framework.
The algorithms are implemented in the \texttt{JAX}~\cite{jax2018github} ecosystem, where the diffusion model relies on \texttt{Flax}~\cite{flax2020github} and FEA relies on \texttt{JAX-FEM}~\cite{xue2023jax}.

\subsection{Conceptual example}
\label{method:example}
To illustrate the proposed method, we consider a conceptual example involving a 1D Gaussian mixture model (GMM). 
The probability density function of a GMM is defined as follows:
\begin{align}
p(x) = \sum_{i=1}^n \pi_i \mathcal{N}(x; \mu_i, \sigma_i^2),
\end{align}
where $n$ denotes the number of Gaussian components, $\pi_i$ represents the mixture weight of component $i$ satisfying $\sum_{i=1}^n \pi_i = 1$, and $\mathcal{N}(x; \mu_i, \sigma_i^2)$ is a Gaussian distribution with mean $\mu_i$ and variance $\sigma_i^2$. 
In this example, we construct a two-component mixture with $n=2$ and parameters $\pi = [0.5, 0.5]$, $\mu = [-2.5, 2.5]$, and $\sigma = [0.5, 0.5]$, resulting in a symmetric bimodal distribution.

\begin{figure}[H] \centering
    {\includegraphics[width=0.8\textwidth]{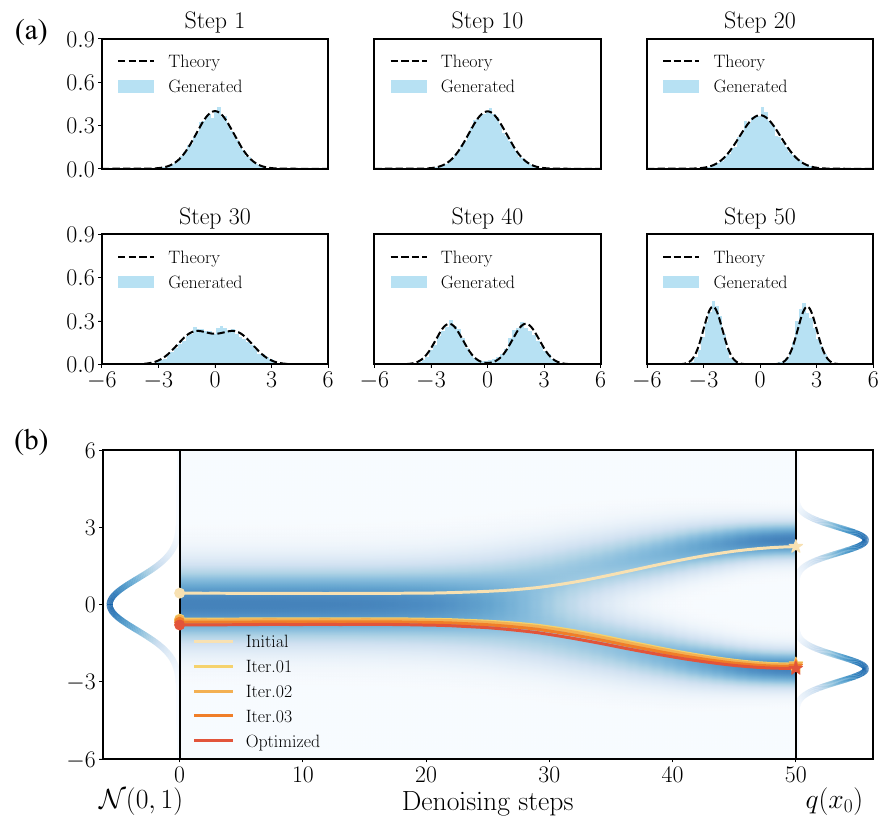}}
    \caption{Design results of the conceptual Gaussian mixture example. (a) The reverse denoising process of the trained diffusion model. (b) The sampling trajectories during the optimization process to transition a sample from the positive mode to the negative mode.
    } \label{fig:gmm}
\end{figure}

First, we use this GMM as the target data distribution $q(x_0)$ and draw $20000$ samples to train a 1D unconditional denoising diffusion model. 
The denoising neural network consists of three fully connected layers with hidden dimensions of $64$ and Swish activation functions, where the timestep is encoded using sinusoidal positional embeddings and integrated through additive connections. 
The forward diffusion process employs $T=1000$ time steps with a linear variance schedule $\beta_t$ ranging from $10^{-4}$ to $0.02$. 
The model is trained for $100$ epochs using the Adam optimizer with a learning rate of $10^{-4}$ and a batch size of $128$. The reverse denoising process follows Eq.~(\ref{eq:ddim}) with $\eta=0$ and $50$ time steps.

Fig.~\ref{fig:gmm} (a) illustrates the reverse sampling process of the trained diffusion model, where the denoising samples $\boldsymbol{x}_t$ at different steps are shown as light blue histograms and the theoretical distribution derived from the forward process is represented by the black dashed curve. 
It can be observed that the denoising samples at each time step closely match the theoretical distribution, and the sample distribution gradually transitions from a standard normal distribution to the desired bimodal distribution, validating the effectiveness of the diffusion model.

Then we consider a simple design task of finding the optimal noise input to guide the generated sample toward a target value $x_{\mathrm{obj}}$, with the loss function defined as
\begin{align}
  \mathcal{L}(\boldsymbol{w}) = \left[\boldsymbol{x}_0\left(\boldsymbol{w}\right) - x_{\mathrm{obj}}\right]^2.
\end{align}
Specifically, we optimize the noise input to transition a sample from the positive mode (near $2.5$) to the negative mode (near $-2.5$) using the gradient-based BFGS algorithm~\cite{nocedal2006numerical}. 
Fig.~\ref{fig:gmm} (b) visualizes the reverse denoising process for the optimization iterations, where different colored lines represent different iterations.
The left panel shows the standard normal distribution, the right panel displays the bimodal GMM distribution, and the middle panel illustrates the denoising trajectory across all timesteps. 
The background colormap represents the theoretical probability distribution at each denoising step, with darker blue indicating higher density regions. 

We observe that all the trajectories traverse high-probability regions at each denoising step, further validating the effectiveness of the trained diffusion model. 
The optimization algorithm progressively modifies the noise inputs across iterations. 
By the first iteration, the trajectory already bends toward the target mode, and by the fourth iteration, the optimized noise input successfully steers the sample to precisely reach the desired target value of $-2.5$ at the negative mode, with a loss below $10^{-3}$. 
This conceptual example demonstrates the core mechanism of our approach: by optimizing the noise input, we can control the reverse denoising trajectory to guide the generated sample toward the desired target. 

\section{Results and discussion}
\label{sec:examples}
This section presents the results of applying the proposed inverse design framework to various microstructure design tasks. We first describe the dataset and denoising neural network, as well as the projection and optimization settings used in the examples. Then, we demonstrate the effectiveness of the proposed method in designing microstructures with tailored homogenized, hyperelastic, and elasto-plastic properties. Finally, we discuss the advantages and limitations of the proposed approach as well as potential future directions.

\subsection{Preparation}
\subsubsection{Dataset and neural network}
\label{examples:pre:nn}
In this study, we utilize two representative types of image data to train the diffusion models: grayscale images of handwritten digits from the MNIST dataset~\cite{lecun2002gradient}, and binary images of 2D orthotropic metamaterial microstructures~\cite{Wang2021}. The MNIST dataset consists of 60000 training images of handwritten digits (0-9) with a resolution of $28\times28$ pixels, which is widely used in the machine learning community for benchmarking image generation and classification tasks. The metamaterial dataset contains 90245 binary orthotropic microstructures with a resolution of $50\times50$ pixels. To facilitate the optimization process, we convert the original binary microstructures into grayscale images by first adding Gaussian noise with zero mean and standard deviation of $0.01$, and then applying a Gaussian filter with standard deviation of $1.2$ to smooth the noisy images. It will be demonstrated in Section~\ref{results:homogenization} that the introduction of the intermediate grayscale phase can avoid the loss of gradient information during optimization while maintaining the overall geometric features of the microstructures. Both datasets are rescaled to the range of $[-1, 1]$ and resized to $32\times32$ pixels for training the diffusion models.

The denoising diffusion neural network adopts a U-Net architecture~\cite{ronneberger2015u} with a hierarchical encoder-decoder structure spanning four resolution levels ($32\times32$, $16\times16$, $8\times8$, and $4\times4$). 
Residual blocks~\cite{he2016deep} are incorporated in both the encoder and decoder paths.
Multi-head self-attention mechanisms~\cite{vaswani2017attention} are applied at the intermediate resolution level to capture spatial dependencies and global structural patterns, allowing the model to learn non-local correlations.
Time step information is encoded through sinusoidal positional embeddings, which are subsequently projected by a two-layer feedforward network with GELU activations and integrated into each residual block. 
The forward diffusion process employs $T=1000$ time steps with a linear variance schedule $\beta_t$ ranging from $10^{-4}$ to $0.02$.
The models are trained on the above datasets for $100$ epochs with a batch size of $128$, using the AdamW optimizer~\cite{loshchilov2017decoupled} and a warmup-cosine learning rate schedule.
The reverse denoising process follows Eq.~(\ref{eq:ddim}) with $\eta=0$ for deterministic generation and employs $10$ time steps for efficiency.

\subsubsection{Projection and optimization}
The trained diffusion models can take random noise inputs and generate grayscale images with similar stylistic features as those in the training datasets.
Each image is interpreted as a $32\times 32$ finite element mesh over the unit square $(0,1)\times(0,1)$, where each pixel corresponds to one element. 
The material properties of each element are then assigned via a linear interpolation scheme
\begin{equation}
  \Theta(\tilde{\boldsymbol{x}}) = \Theta_{0} + (\Theta_{1} - \Theta_{0}) \tilde{\boldsymbol{x}},
\end{equation}
where $\Theta_{0}$ and $\Theta_{1}$ are the properties of the two base materials. $\tilde{\boldsymbol{x}}$ denotes the material density field obtained from the generated image $\boldsymbol{x}_0$ via a projection function. Specifically, we set $\tilde{\boldsymbol{x}} = 1 - \mathcal{P}(\boldsymbol{x}_0)$ for the MNIST dataset and $\tilde{\boldsymbol{x}} = \mathcal{P}(\boldsymbol{x}_0)$ for the metamaterial dataset. The projection function $\mathcal{P}(\boldsymbol{x})$ is defined as
\begin{equation}
  \mathcal{P}(\boldsymbol{x}) = \frac{1}{2} \left[ \tanh(\gamma\boldsymbol{x}) + 1 \right],
\end{equation}
where $\gamma$ is a hyperparameter that controls the sharpness of the projection. A small $\gamma$ results in a smooth transition between material phases, while a large $\gamma$ produces a more binary distribution.
The optimization is performed using the gradient-based BFGS algorithm~\cite{nocedal2006numerical}. All computations are carried out on an Apple M3 Pro platform with 36 GB of RAM.

Although the existence of the intermediate material phase can facilitate the gradient-based optimization, the final design is expected to be a nearly binary distribution suitable for practical manufacturing. To address this issue, we perform a multi-stage optimization strategy: the optimization starts with a small $\gamma$ to allow for smooth material transitions, and then the optimal design serves as the initial guess for the next stage with an increased $\gamma$. This process is repeated until a sufficiently large $\gamma$ is reached, resulting in a nearly binary material distribution. To evaluate the binary quality of the generated microstructures, we define a binarization metric as follows:
\begin{equation}
  \mathcal{B}(\tau) = \frac{1}{n} \sum_{i=1}^n \mathbb{I}\!\left(\tilde{x}_i \le \tau \,\lor\, \tilde{x}_i \ge 1-\tau \right),
\end{equation}
where $\mathbb{I}(\cdot)$ is the indicator function, $n$ is the total number of pixels, $\tilde{x}_i$ is the projected density of pixel $i$, and $\tau$ is a small threshold chosen as $10^{-3}$ in this work. A higher value of $\mathcal{B}$ indicates a more binary design. As we will demonstrate in the following examples, this multi-stage strategy can effectively guide the optimization towards manufacturable designs with desired mechanical properties.

\subsection{Tailored homogenized properties}
\label{results:homogenization}
\begin{figure}[H] \centering
    {\includegraphics[width=0.9\textwidth]{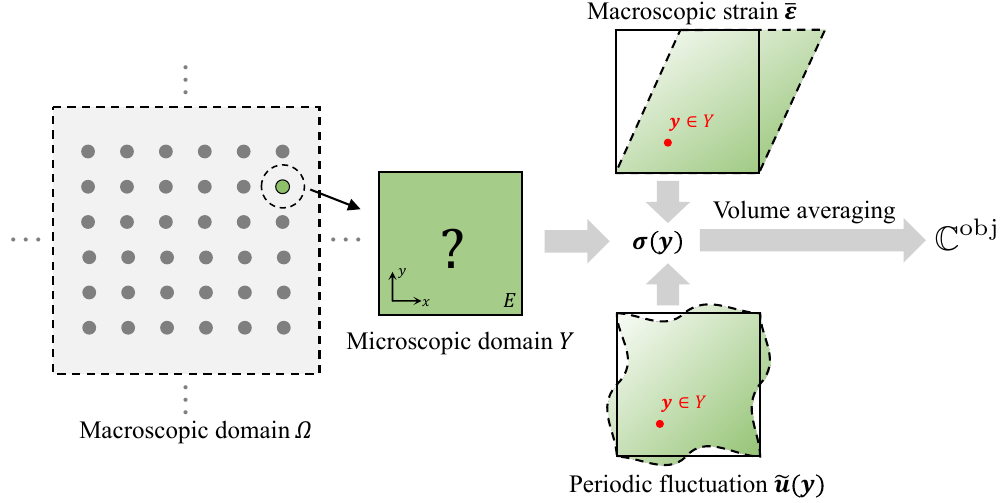}}
    \caption{Schematic of the model setup for the microstructure design example with tailored homogenized properties.
    }\label{fig:homo_model}
\end{figure}

In the first example, we apply the proposed method to the design of microstructures with tailored homogenized properties, providing a fundamental building block for advanced composite materials.
The generated microstructure is modeled as a representative volume element (RVE) $Y$ with periodic boundary conditions for the macroscopic domain $\Omega$, as illustrated in Fig.~\ref{fig:homo_model}. The computational homogenization is performed by applying a set of prescribed macroscopic strain load cases $\bar{\boldsymbol{\varepsilon}}$ to the RVE and solving the following boundary value problem:
\begin{subequations}
\begin{align}
  -\nabla \cdot \boldsymbol{\sigma} = \mathbf{0}\quad &\text{in} \ Y,\\
  \boldsymbol{\sigma}\left(\boldsymbol{y}\right) = \mathbb{C}\left(\boldsymbol{y}\right) : \boldsymbol{\varepsilon}\left(\boldsymbol{y}\right)\quad &\text{for}\ \boldsymbol{y}\in Y,\\
  \boldsymbol{\varepsilon}\left(\boldsymbol{y}\right) = \bar{\boldsymbol{\varepsilon}} + \nabla^s \tilde{\boldsymbol{u}} \quad &\text{in} \ Y,\\
  \tilde{\boldsymbol{u}}\quad &\text{is} \ Y\text{-periodic},\\
  \boldsymbol{\sigma} \cdot \boldsymbol{n}=\boldsymbol{t} \quad &\text{is } Y\text{-antiperiodic on } \partial Y,
\end{align}
\end{subequations}
where $\boldsymbol{y}$ is the microscopic coordinate, $\boldsymbol{\sigma}$ is the microscopic Cauchy stress, $\boldsymbol{\varepsilon}$ is the microscopic strain, $\mathbb{C}$ is the local elasticity tensor assigned based on the generated microstructure, $\tilde{\boldsymbol{u}}$ is the periodic fluctuation displacement field to be solved, $\boldsymbol{t}$ is the traction vector on the RVE boundary $\partial Y$, and $\boldsymbol{n}$ is the outward normal vector on $\partial Y$. 
The macroscopic stress $\bar{\boldsymbol{\sigma}}$ is computed as the volume average of $\boldsymbol{\sigma}$ over the RVE
\begin{align}
  \bar{\boldsymbol{\sigma}} = \frac{1}{|Y|} \int_{Y} \boldsymbol{\sigma}(\boldsymbol{y}) \, \mathrm{d}\boldsymbol{y},
\end{align}
where $|Y|$ is the volume of the RVE. Three independent macroscopic strain load cases, including uniaxial tension along the $x$- and $y$-directions as well as pure shear, are applied to the RVE to compute the corresponding macroscopic stresses, and the effective elasticity tensor $\mathbb{C}^{\mathrm{hom}}$ can be derived from the macroscopic stress-strain relation
\begin{align}
  \bar{\boldsymbol{\sigma}} = \mathbb{C}^{\mathrm{hom}} : \bar{\boldsymbol{\varepsilon}}.
\end{align}

The effective elasticity tensor $\mathbb{C}^{\mathrm{hom}}$ characterizes the homogenized elastic response of the composite material at the macroscopic scale. Thus, the ability to achieve a target $\mathbb{C}^{\mathrm{obj}}$ through microstructure design is highly desirable for advanced material applications. In this example, we set the loss function as the squared error between the components of the generated and target effective elasticity tensor
\begin{align}
  \mathcal{L}\left(\boldsymbol{w}\right) = \sum_{i,j=1}^{3} \left[ \mathbb{C}^{\mathrm{hom}}_{ij}\left(\boldsymbol{w}\right) - \mathbb{C}^{\mathrm{obj}}_{ij} \right]^2.
\end{align}

We consider two different base materials for the microstructure design: a soft material with Young's modulus $E_0 = 1$ and a stiff material with Young's modulus $E_1 = 100$, both having a Poisson's ratio of $\nu = 0.3$. We first utilize the MNIST-based diffusion model to perform the microstructure design tasks. We consider three groups with target $\mathbb{C}^{\mathrm{hom}}$ of different symmetries~\cite{desmorat2019space}, including the biclinic, orthotropic, and tetragonal. The target effective elasticity tensors for each group are set as follows:
\begin{align}
  \mathbb{C}^{\mathrm{obj}}_{\text{bic}} = \begin{bmatrix}
    50 & 12 & 0 \\
    12 & 60 & -3 \\
    0 & -3 & 15
  \end{bmatrix},\qquad
  &\mathbb{C}^{\mathrm{obj}}_{\text{ort}} = \begin{bmatrix}
    50 & 12 & 0 \\
    12 & 60 & 0 \\
    0 & 0 & 15
  \end{bmatrix},\qquad 
  \mathbb{C}^{\mathrm{obj}}_{\text{tet}} = \begin{bmatrix}
    60 & 12 & 0 \\
    12 & 60 & 0 \\
    0 & 0 & 15
  \end{bmatrix}.
\end{align}
Each group is initialized with two different noise inputs and optimized using the multi-stage strategy with $\gamma$ values of $5$, $10$, $20$, and $80$. Each stage is terminated when the loss is less than $10^{-2}$ or the maximum number of iterations ($100$) is reached. 

\begin{figure}[H] \centering
    {\includegraphics[width=1.0\textwidth]{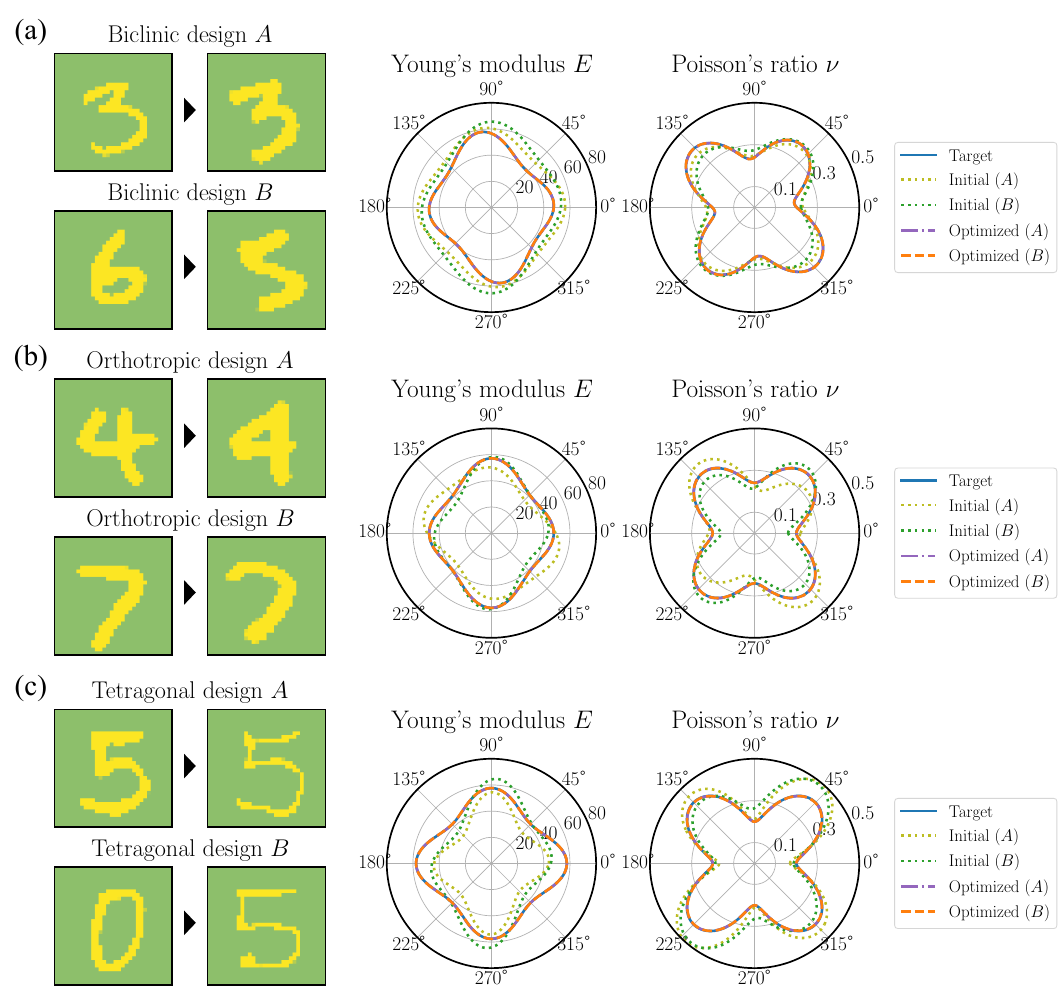}}
    \caption{Design results of the MNIST-based microstructures for achieving target effective elasticity tensors with different symmetries~\cite{desmorat2019space}: (a) Biclinic, (b) Orthotropic, (c) Tetragonal.
    } \label{fig:homo_mnist}
\end{figure}

\begin{figure}[H] \centering
    {\includegraphics[width=0.8\textwidth]{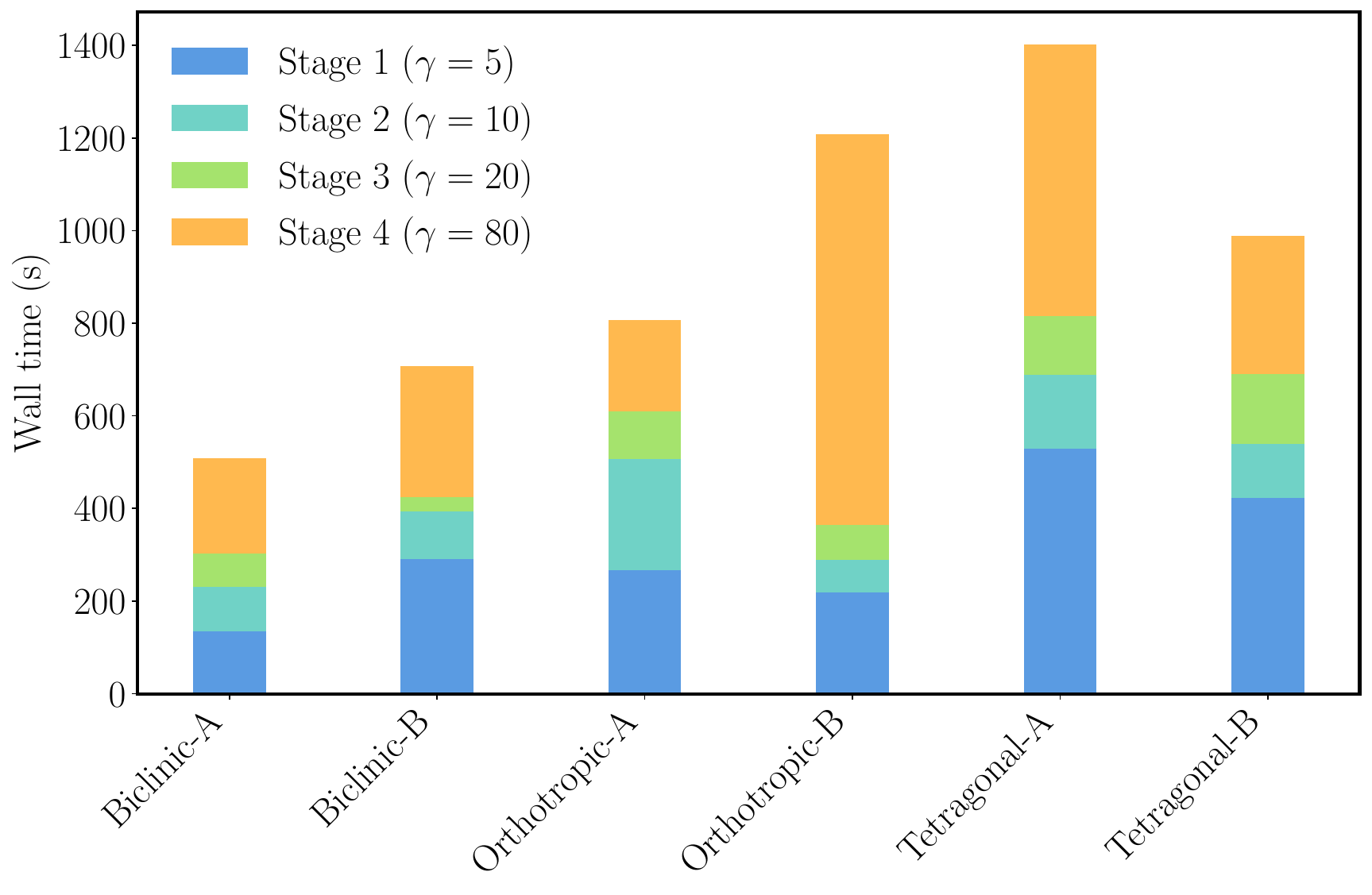}}
    \caption{Wall time for the inverse design of MNIST-based microstructures targeting effective elasticity tensors with different symmetries.
    } \label{fig:homo_mnist_time}
\end{figure}

The design results for each group are shown in Fig.~\ref{fig:homo_mnist}, where the initial and optimized microstructures are presented along with their corresponding Young's modulus and Poisson's ratio in different directions. 
For most cases, the optimized microstructures preserve the overall shape of the initial guess, with local geometric features adjusted to meet the target properties. 
We also observe that some optimized microstructures transition to different topologies compared to the initial guesses, while still maintaining similar geometric features as the handwritten digits. 
For instance, in the Biclinic-B group, the optimized microstructure transitions from a digit 6-like shape to a digit 5-like shape. 
Although the optimized designs exhibit diverse geometric features, they all achieve the desired target properties, demonstrating the capability of the proposed method to steer the design toward specified objectives while adhering to the learned data distribution. 
All binarization metrics $\mathcal{B}$ for the optimized microstructures are around $99\%$, indicating that the multi-stage optimization strategy effectively generates manufacturable designs. 
The wall time for each design case is shown in Fig.~\ref{fig:homo_mnist_time}, with a range from $509$ s to $1402$ s. 
For the four stages of optimization, the first and last stages generally consume more time due to the exploration from the initial guess and the less effective gradient information in the nearly binary stage with large $\gamma$. 

Following the design examples with the MNIST dataset, the proposed method is further examined using the open dataset containing 2D orthotropic metamaterial microstructures~\cite{Wang2021} to perform inverse design tasks.
The base material and optimization settings remain the same as those in the MNIST-based examples, with four optimization stages using $\gamma$ values of $5$, $10$, $20$, and $100$. 
 The target effective elasticity tensors are defined as follows:
\begin{align}
  \mathbb{C}^{\mathrm{obj}}_{\text{ort}} = \begin{bmatrix}
    50 & 15 & 0 \\
    15 & 70 & 0 \\
    0 & 0 & 15
  \end{bmatrix},\qquad 
  \mathbb{C}^{\mathrm{obj}}_{\text{tet}} = \begin{bmatrix}
    50 & 15 & 0 \\
    15 & 50 & 0 \\
    0 & 0 & 15
  \end{bmatrix}.
\end{align}
with two different noise inputs for each symmetry group.

\begin{figure}[H] \centering
    {\includegraphics[width=1.0\textwidth]{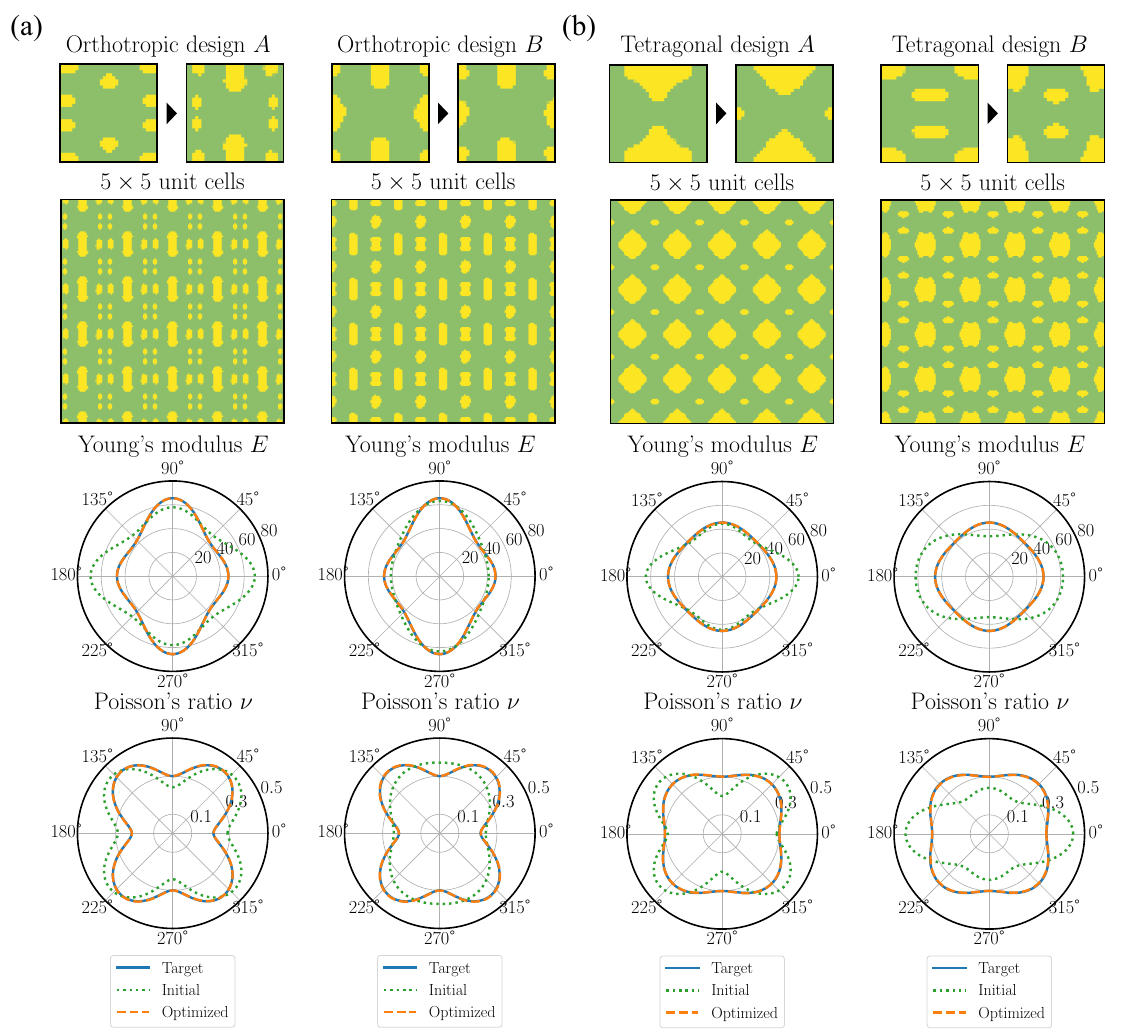}}
    \caption{Design results of the metamaterial-based microstructures for achieving target effective elasticity tensors with different symmetries: (a) Orthotropic and (b) Tetragonal.
    } \label{fig:homo_micro}
\end{figure}

Fig.~\ref{fig:homo_micro} shows the design results for the considered orthotropic and tetragonal symmetries, including the initial and optimized microstructures, the $5\times5$ unit cells of the optimized microstructures, and the corresponding Young's modulus and Poisson's ratio polar plots. 
All the cases achieve a loss below $10^{-2}$ and a binarization metric above $98\%$, indicating the effectiveness of the proposed method in generating manufacturable microstructures with desired homogenized properties. 
Besides, although the training images are strictly symmetric, the grayscale preprocessing, the nature of the diffusion model, and the optimization process can introduce slight asymmetries in the final designs. 
These minor deviations do not significantly impact the overall mechanical performance. 
For orthotropic and tetragonal symmetries, one can select a quarter of the generated image, apply symmetry operations to reconstruct the full microstructure, and then follow the same procedure to obtain strictly symmetric designs.
This issue will be further discussed in Section~\ref{discussion}. 

\begin{figure}[H] \centering
    {\includegraphics[width=0.9\textwidth]{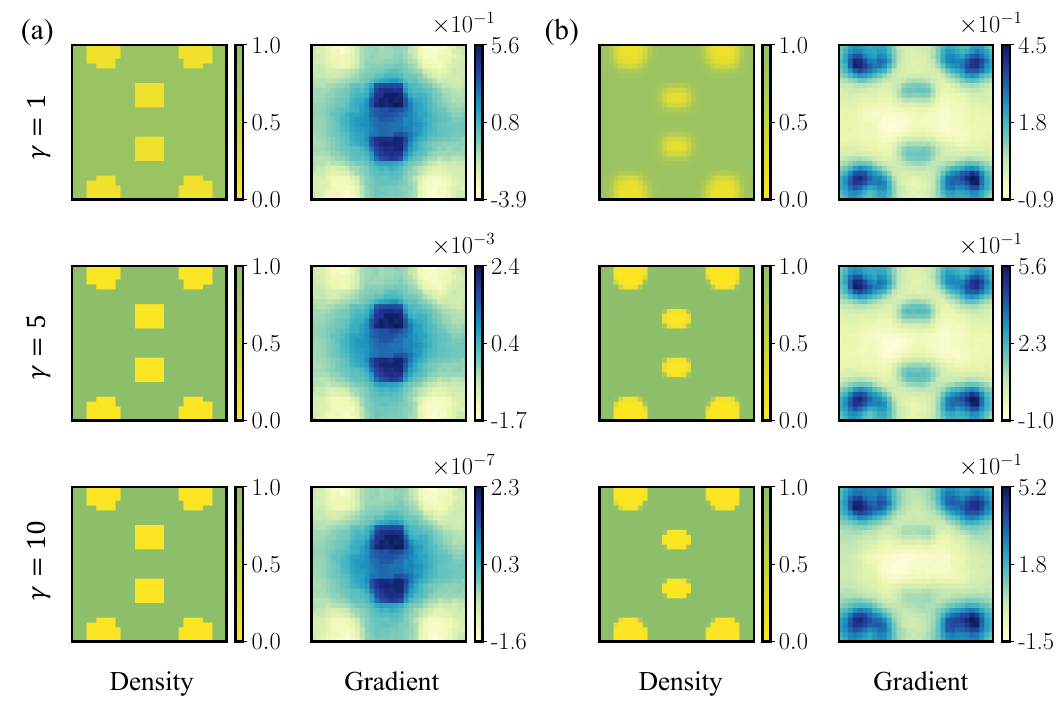}}
    \caption{Comparison under different slope $\gamma$. Projected density field (left column) and the gradient of $\mathbb{C}^{\mathrm{hom}}_{11}$ with respect to the noise input (right column) for (a) binary-based and (b) grayscale-based diffusion models.
    } \label{fig:homo_compare}
\end{figure}

As mentioned in Section~\ref{examples:pre:nn}, we preprocess the original binary microstructure images into grayscale ones to facilitate gradient-based optimization. 
To illustrate the reason behind this, we conduct a comparative study between the grayscale-based and binary-based diffusion models. 
Specifically, we adopt the same settings as those in Section~\ref{examples:pre:nn} to train another diffusion model using the original binary data. 
Fig.~\ref{fig:homo_compare} shows comparison results generated from both the grayscale-based and binary-based diffusion models, including the projected density fields with three $\gamma$ values of $1$, $5$, and $10$, and the corresponding gradients of the component $C_{11}^{\mathrm{hom}}$ with respect to the noise input. 
We observe that the binary-based diffusion model generates nearly $0$-$1$ density fields, and the gradient magnitude gradually decreases as $\gamma$ increases. 
On the other hand, the grayscale-based model produces relatively smooth density fields with intermediate values and maintains significant gradient magnitudes even with large $\gamma$.
This is because the binary-based diffusion model tends to produce pixel values that lie near the boundary of the training data ($-1$ and $1$), resulting in vanishing gradients when the projection function becomes steep with large $\gamma$. 
In contrast, the grayscale-based model allows for the generation of intermediate pixel values, enabling non-zero gradients even with a sharp projection. 
Besides, the generated grayscale microstructures can still capture the overall geometric features of the original binary designs, and the final optimized results can achieve high binarization metrics through the multi-stage strategy.
Therefore, the grayscale-based diffusion model is adopted for the gradient-based optimization tasks in this work.

\subsection{Tailored hyperelastic properties}
\label{results:hyper}
Apart from achieving specific homogenized properties, controlling the mechanical response of microstructures is essential for attaining desired functional performance.
In particular, tailoring the hyperelastic behavior of microstructures is crucial for applications such as biomedical devices~\cite{kazim2025mechanical} and soft robotics~\cite{zheng2025metamaterial}.
In this section, we consider the design of microstructures with tailored hyperelastic behaviors. 
The governing equations for the deformation of a hyperelastic body $\Omega$ are given by
\begin{subequations}
\begin{align}
  -\nabla \cdot \boldsymbol{P} = \mathbf{0}\quad &\text{in} \ \Omega,\\
  \boldsymbol{u} = \bar{\boldsymbol{u}}\quad &\text{on} \ \Gamma_D,\\
  \boldsymbol{P} \cdot \boldsymbol{N} = \bar{\boldsymbol{t}}\quad &\text{on} \ \Gamma_N,
\end{align}
\end{subequations}
where $\boldsymbol{u}$ is the displacement field to be solved. $\boldsymbol{P}=\frac{\partial W}{\partial \boldsymbol{F}}$ is the first Piola-Kirchhoff stress, $W$ is the strain energy density function, and $\boldsymbol{F} = \mathbf{I} + \nabla \boldsymbol{u}$ is the deformation gradient. $\bar{\boldsymbol{u}}$ is the prescribed displacement on the Dirichlet boundary $\Gamma_D$, and $\bar{\boldsymbol{t}}$ is the prescribed traction on the Neumann boundary $\Gamma_N$ with unit outward normal $\boldsymbol{N}$. The base material is modeled using the Neo-Hookean hyperelastic model, with the $W$ defined as
\begin{equation}
  W = \frac{G}{2}(J^{-2/3}I_1 - 3) + \frac{K}{2}(J - 1)^2
\end{equation}
where $G=\frac{E}{2(1+\nu)}$ is the shear modulus, $K=\frac{E}{3(1-2\nu)}$ is the bulk modulus, $E$ is the Young's modulus, $\nu$ is the Poisson's ratio, $I_1 = \text{tr}(\boldsymbol{C})$ is the first invariant of the right Cauchy-Green deformation tensor $\boldsymbol{C} = \boldsymbol{F}^\top \boldsymbol{F}$, and $J = \det(\boldsymbol{F})$ is the determinant of $\boldsymbol{F}$. 

\begin{figure}[H] \centering
    {\includegraphics[width=0.8\textwidth]{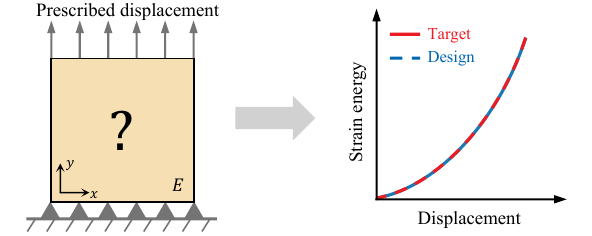}}
    \caption{Schematic of the model setup for the microstructure design example with tailored hyperelastic properties.
    } \label{fig:hyper_model}
\end{figure}

As illustrated in Fig.~\ref{fig:hyper_model}, we consider the design of microstructures with tailored strain energy curves under uniaxial tension. The two base materials are set as a stiff material with Young's modulus $E_1 = 100$ and a soft material with Young's modulus $E_0 = 1$, both having a Poisson's ratio of $\nu = 0.3$. A uniform displacement loading from $0.1$ to $0.5$ is applied on the top boundary of the computational domain, while the bottom boundary is fixed. To represent the target hyperelastic behaviors, we first approximate the strain energy curve function of the microstructures with full base materials using third-order polynomials: $P_{1}=-19.880x^3+51.245x^2+0.472x$ and $P_{0}=-0.199x^3+0.512x^2+0.005x$ for the stiff and soft base materials, respectively. Then, we define the target strain energy curve as a linear interpolation between the two base material curves
\begin{align}
  P^{*} = (1-\alpha)P_{1} + \alpha P_{0},
\end{align}
where $\alpha \in [0, 1]$ is the interpolation parameter controls the desired hyperelastic behavior. The loss function is defined as the mean squared error across all loading steps between the computed and target strain energy 
\begin{align}
\mathcal{L}\left(\boldsymbol{w}\right) = \frac{1}{N} \sum_{i=1}^N \left[ P_i\left(\boldsymbol{w}\right) - P^{*}\left(u_i\right) \right]^2
\end{align}
where $N$ is the total number of loading steps, $P_i$ is the strain energy of the generated microstructure under the $i$-th applied displacement $u_i$, and $P^{*}(u_i)$ is the corresponding target value. The multi-stage optimization is performed with $\gamma$ values of $2$, $5$, $10$, $20$, and $100$. Each stage is terminated when the loss is less than $10^{-3}$ or the maximum number of iterations ($30$) is reached.

\begin{figure}[H]
    \centering
    {\includegraphics[width=1.0\textwidth]{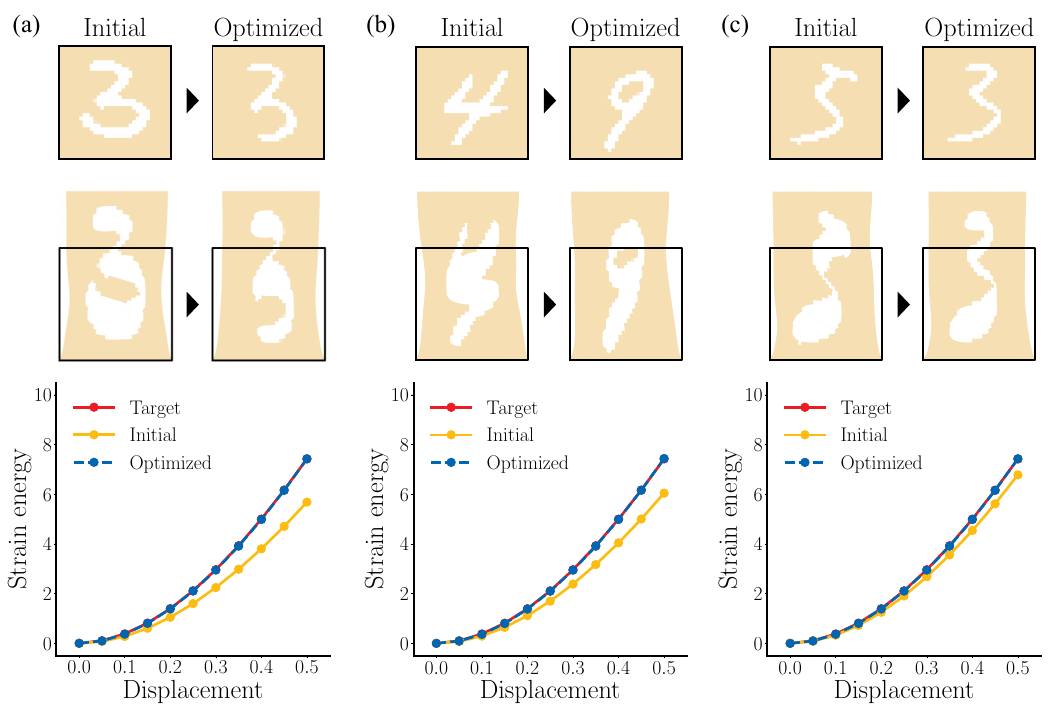}}
    \caption{Design results for hyperelastic microstructures with the same strain energy curve and different initial designs: (a) digit 3; (b) digit 4; (c) digit 5. From top to bottom: undeformed configuration, deformed configuration under the maximum displacement loading, and the corresponding strain energy curves.
    }
    \label{fig:hyper_ini}
\end{figure}

Fig.~\ref{fig:hyper_ini} shows the design results for hyperelastic microstructures with the same strain energy curve $\alpha=0.3$ and different initial designs, including digits 3, 4, and 5. 
The corresponding strain energy curves of the initial and optimized microstructures compared with the target curves are shown at the bottom. 
Although the initial microstructures exhibit significantly different hyperelastic behaviors, the proposed method can effectively optimize them to achieve the desired target properties while maintaining handwritten digit-like styles.
The optimized microstructure starting from digit 3 retains the overall shape of digit 3 but with a reduced scale to accommodate larger stiffness requirements. In contrast, the optimizations starting from digits 4 and 5 converge to different shapes resembling digits 9 and 3, respectively, while still achieving the desired hyperelastic response. 
All optimized microstructures achieve binarization metrics $\mathcal{B}$ above $99\%$, indicating that the multi-stage optimization strategy successfully generates manufacturable designs.

\begin{figure}[H] \centering
    {\includegraphics[width=1.0\textwidth]{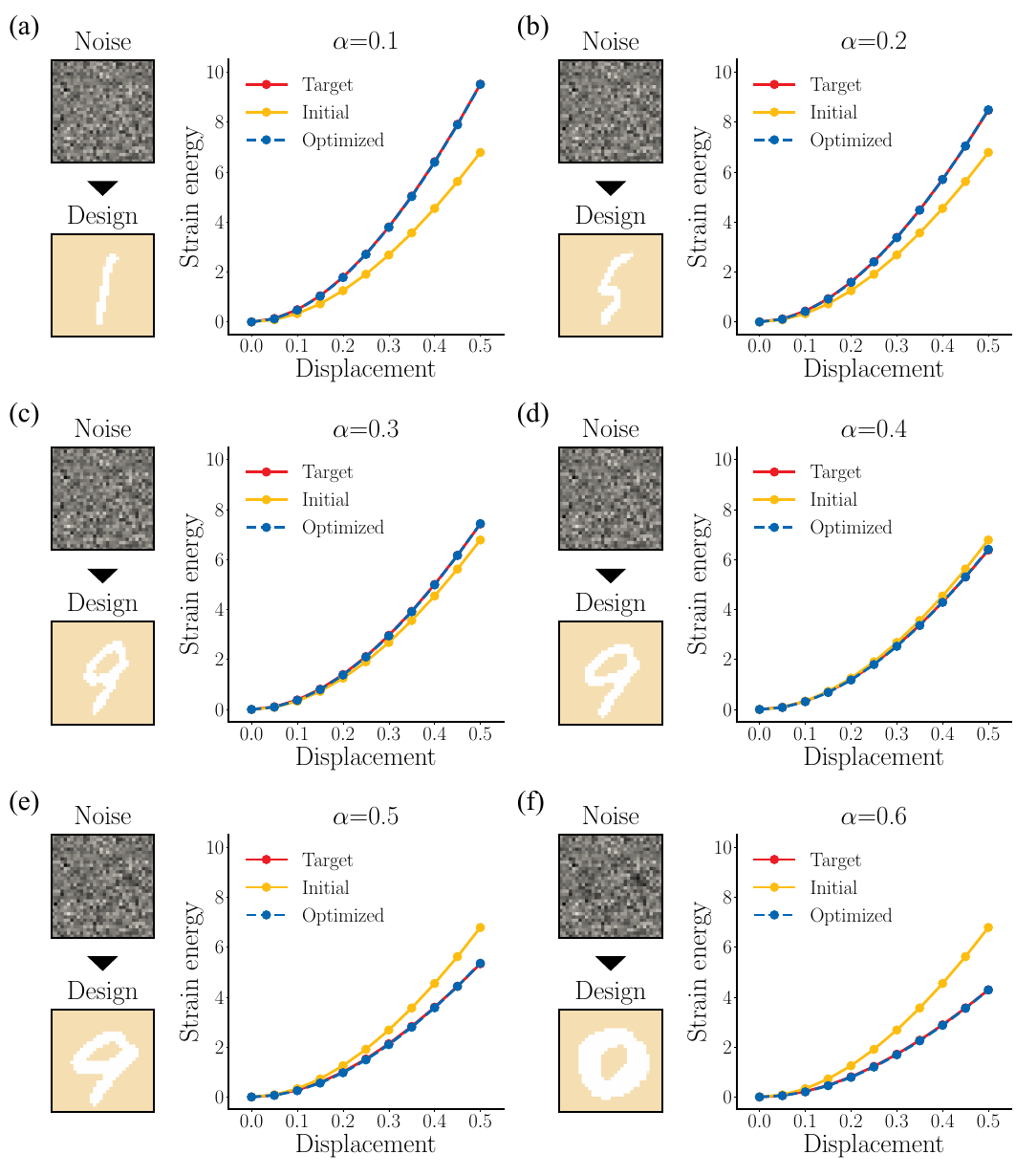}}
    \caption{Design results for hyperelastic microstructures with the same initial design and different target strain energy curves characterized by $\alpha$ values. $(a)$ to $(f)$ correspond to $\alpha=0.1$, $0.2$, $0.3$, $0.4$, $0.5$, and $0.6$, respectively. 
    } 
    \label{fig:hyper_alphas}
\end{figure}

Then we investigate the influence of different target hyperelastic behaviors on the optimized microstructures with the same initial design.
Fig.~\ref{fig:hyper_alphas} presents the initial and optimized microstructures as well as the corresponding strain energy curves, with the interpolation parameter $\alpha$ ranging from 0.1 to 0.6. 
The results demonstrate the capability of the proposed method to generate microstructures that closely match a variety of target hyperelastic behaviors. 
All optimized microstructures achieve binarization metrics above $99\%$, indicating the effectiveness of the multi-stage optimization strategy.

Besides, we can observe a gradual transition in the optimized microstructures to accommodate the changing hyperelastic response. 
When $\alpha$ is small, the strain energy curve is closer to that of the stiff base material, resulting in slim shapes like digits 1 and 5.
As $\alpha$ increases, the optimized microstructures incorporate more soft material regions to achieve lower stiffness, gradually resembling digit 9 and eventually 0. 
For $\alpha$ larger than $0.6$, the optimized microstructures tend to converge to similar designs as $\alpha=0.6$ but with more soft material regions. Meanwhile, the optimized noise input tends to deviate from the isotropic Gaussian distribution. These observations indicate the limitations of selected base material properties in achieving such extremely soft hyperelastic properties while preserving the clear geometric features encoded in the diffusion model. One can address this issue by choosing other combinations of base materials to achieve the desired performance and style. This phenomenon will be further discussed in Section~\ref{discussion}.

\subsection{Tailored elasto-plastic properties}
\label{results:plast}
In addition to hyperelastic responses, elasto-plastic behaviors represent another important class of nonlinear material responses prevalent in engineering applications. This section considers the inverse design of microstructures with tailored elasto-plastic properties. The governing equations for the elasto-plastic deformation of materials are given by
\begin{subequations}
\begin{align}
  -\nabla \cdot \boldsymbol{\sigma}^k = \boldsymbol{0}\quad &\text{in} \ \Omega,\\
  \boldsymbol{u}^k = \bar{\boldsymbol{u}}\quad &\text{on} \ \Gamma_D,\\
  \boldsymbol{\sigma}^k\cdot\boldsymbol{n} = \bar{\boldsymbol{t}}\quad &\text{on} \ \Gamma_N,
\end{align}
\end{subequations}
where $\boldsymbol{\sigma}^k = \boldsymbol{f}(\boldsymbol{u}^k, \boldsymbol{\varepsilon}^{k-1}, \boldsymbol{\sigma}^{k-1})$ is the updated Cauchy stress at time step $k$, obtained from the stress update function $\boldsymbol{f}$ that depends on the current displacement $\boldsymbol{u}^k$ as well as the total strain $\boldsymbol{\varepsilon}^{k-1}$ and stress $\boldsymbol{\sigma}^{k-1}$ from the previous time step. The notations related to boundary conditions are the same as those in previous sections. Here we adopt the perfect J2-plasticity model~\cite{simo1998computational} with small strain assumption to describe the elasto-plastic behavior of the base material. The stress update function $\boldsymbol{f}$ is defined as follows:
\begin{subequations}
\begin{align}
  \boldsymbol{f}(\boldsymbol{u}^k, \boldsymbol{\varepsilon}^{k-1}, \boldsymbol{\sigma}^{k-1}) &= \boldsymbol{\sigma}_{\text{trial}}-\frac{\boldsymbol{s}_{\text{trial}}}{s_{\text{trial}}}\langle f_{\text{yield}} \rangle_+,\\
  \boldsymbol{\sigma}_{\text{trial}} &= \boldsymbol{\sigma}^{k-1} + \mathbb{C} : ( \nabla^s \boldsymbol{u}^k - \boldsymbol{\varepsilon}^{k-1}),\\
  f_{\text{yield}} &= s_{\text{trial}} - \sigma_0,
\end{align}
\end{subequations}
where $\boldsymbol{\sigma}_{\text{trial}}$ is the trial stress, $\mathbb{C}$ is the elasticity tensor of the base material, $\boldsymbol{s}_{\text{trial}}$ is the deviatoric part of $\boldsymbol{\sigma}_{\text{trial}}$, $s_{\text{trial}}=\sqrt{\frac{3}{2}\boldsymbol{s}_{\text{trial}}:\boldsymbol{s}_{\text{trial}}}$ is the equivalent von Mises stress, $\sigma_0$ is the yield stress of the base material, and $\langle \cdot \rangle_+$ is the ramp function.

\begin{figure}[H] \centering
    {\includegraphics[width=0.8\textwidth]{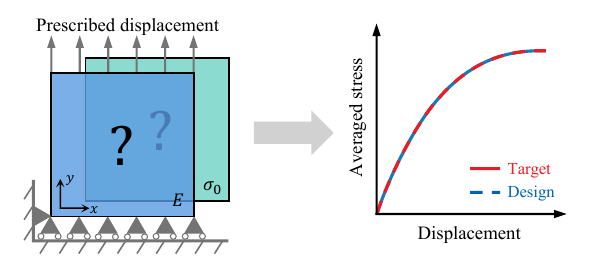}}
    \caption{Schematic of the model setup for the microstructure design example with tailored elasto-plastic properties.
    } 
    \label{fig:plast_model}
\end{figure}

As presented in Fig.~\ref{fig:plast_model}, we consider the design task of microstructures with target stress-displacement curves under uniaxial tension. Two base materials are considered: a stiff material with Young's modulus $E_1 = 10^5$ and yield stress $\sigma_{0}^{(1)} = 300$, and a soft material with Young's modulus $E_0 = 10^3$ and yield stress $\sigma_{0}^{(0)} = 30$. Both materials have a Poisson's ratio of $\nu = 0.3$. The top boundary of the computational domain is subjected to a uniform displacement loading in the $y$ direction, ranging from $5\times10^{-4}$ to $10^{-2}$. 
The bottom boundary is fixed in the $y$ direction, while the left-bottom corner is also fixed in the $x$ direction. 
The design objective function is formulated as follows:
\begin{align}
\mathcal{L}\left(\boldsymbol{w}\right) = \frac{1}{N} \sum_{i=1}^N \left[ \bar{\sigma}_i\left(\boldsymbol{w}\right) - \bar{\sigma}_i^{\mathrm{obj}} \right]^2
\end{align}
where $N$ is the total number of displacement loading steps. $\bar{\sigma}_i$ represents the averaged stress in the $y$ direction of the generated microstructure at the displacement loading $u_i$, and $\bar{\sigma}_i^{\mathrm{obj}}$ represents the corresponding target value. 

\begin{figure}[H] \centering
    {\includegraphics[width=0.9\textwidth]{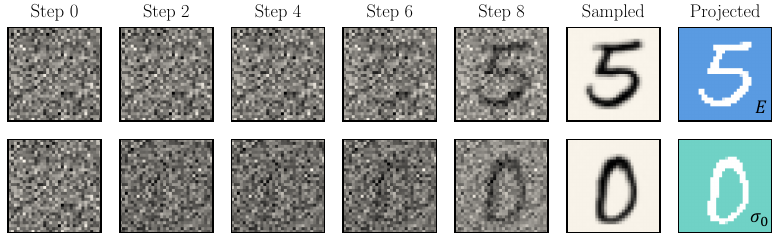}}
    \caption{Reverse denoising process and projected microstructures
    of two randomly sampled noise inputs for the elasto-plastic design example: Young's modulus field (top); yield stress field (bottom).
    } 
    \label{fig:plast_sample}
\end{figure}

\begin{figure}[H]
    \centering
    {\includegraphics[width=1.0\textwidth]{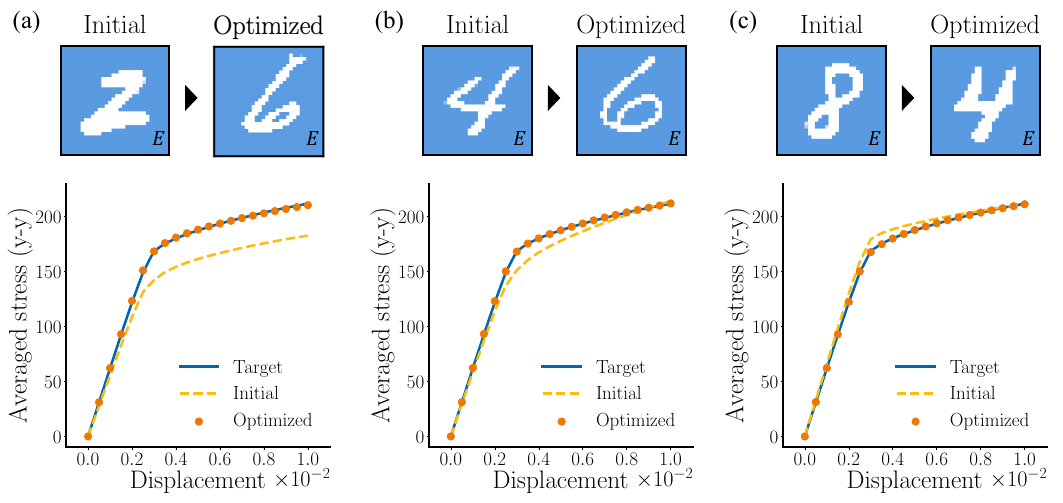}}
    \caption{Elasto-plastic microstructure designs matching the same target stress-displacement curve. Young's modulus fields are optimized starting from different initial designs: (a) digit 2; (b) digit 4; (c) digit 8.}
    \label{fig:plast_E}
\end{figure}

We first investigate the design of a single material property field, either Young's modulus $E$ or yield stress $\sigma_0$, while keeping the other property field fixed. As shown in Fig.~\ref{fig:plast_sample}, we randomly sample two different digits from the trained MNIST-based diffusion model to represent the Young's modulus (top) and yield stress fields (bottom). The projected microstructures with $\gamma=100$ are shown in the last column of Fig.~\ref{fig:plast_sample}, which are used to compute the stress-displacement curve as the target. When designing the Young's modulus field, the yield stress field is fixed as the one in Fig.~\ref{fig:plast_sample} and vice versa. The multi-stage optimization is performed with $\gamma$ values of $5$, $10$, $20$, and $100$, together with a loss tolerance of $0.2$ and a maximum iteration number of $50$ for each stage.

Fig.~\ref{fig:plast_E} shows the design results for the Young's modulus field with three different initial microstructures, including digits 2, 4, and 8. 
The optimized microstructures of 2 and 4 converge to similar designs resembling the digit 6, and that of 8 converges to a design similar to the digit 4. 
As shown in the bottom of Fig.~\ref{fig:plast_E}, the optimized microstructures successfully achieve the target elasto-plastic behaviors, regardless of the apparently different initial designs. 

\begin{figure}[H]
    \centering
    {\includegraphics[width=1.0\textwidth]{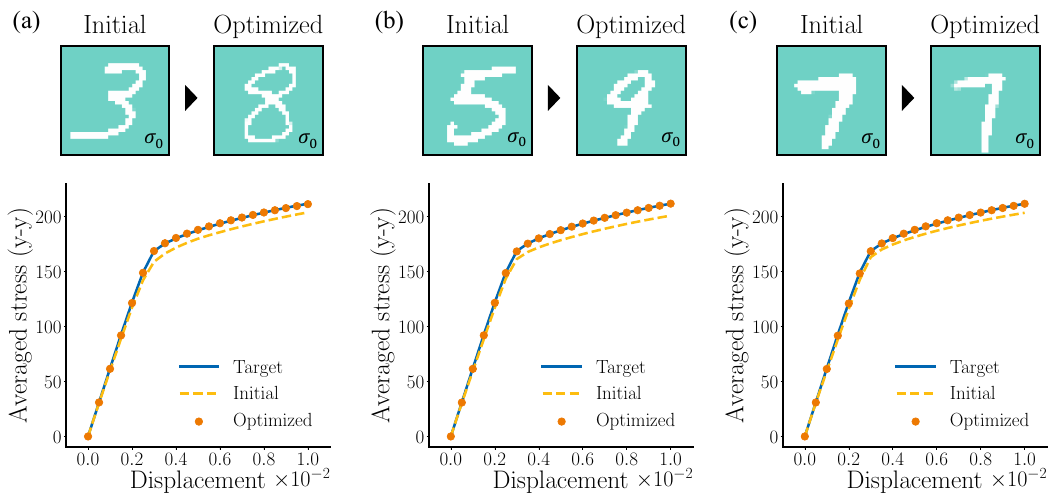}}
    \caption{Elasto-plastic microstructure designs matching the same target stress-displacement curve. Yield stress fields are optimized starting from different initial designs: (a) digit 3; (b) digit 5; (c) digit 7.}
    \label{fig:plast_sig0}
\end{figure}

The design results for the yield stress field with different initial microstructures are presented in Fig.~\ref{fig:plast_sig0}. 
Three different initial designs, including digits 3, 5, and 7, are considered. 
The optimized microstructures of 3 and 5 converge to similar designs resembling digits 8 and 9, respectively, while that of 7 converges to a reduced scale of digit 7. 
Although the optimized yield stress fields exhibit different geometric features, they all successfully achieve the target elasto-plastic behaviors, as shown in the bottom of Fig.~\ref{fig:plast_sig0}.

After demonstrating the effectiveness of the proposed method in designing a single material property field, we consider the simultaneous design of both Young's modulus and yield stress fields to achieve tailored elasto-plastic properties. To represent the target elasto-plastic behaviors in a parametric form, we use a piecewise function to describe the stress-displacement curve of the MNIST-based microstructures
\begin{align}
f(x) = 
\begin{cases}
Ex, & x \leq \varepsilon_y \\
\sigma_0 + \sum_{i=1}^{n} A_i \left(1 - e^{-\kappa_i(x - \varepsilon_y)}\right), & x > \varepsilon_y
\end{cases}
\end{align}
where the linear elastic region is characterized by Young's modulus $E$ up to the yield strain $\varepsilon_y$ with yield stress $\sigma_0$, and the post-yield plastic behavior is modeled using a sum of exponential functions with amplitudes $A_i$ and rates $\kappa_i$. The summation of $A_i$ equals to $\sigma_{\infty} - \sigma_0$, where $\sigma_{\infty}$ is the asymptotic stress at large strains. Since $\sigma_0 = E \varepsilon_y$, the target elasto-plastic behavior can be fully defined by the parameters $\{E, \varepsilon_y, \sigma_{\infty}, A_i, \kappa_i\}$. In this example, we set $n=3$ and consider three target stress-displacement curves with different properties, and the corresponding parameters are listed in Table~\ref{tab:plast_params}. The initial noise inputs remain the same as those in Fig.~\ref{fig:plast_sample}, with a digit $5$ for the Young's modulus field and a digit $0$ for the yield stress field. Three $\gamma$ values of $10$, $20$, and $100$ are used for the multi-stage optimization, with a loss tolerance of $0.5$ and a maximum iteration number of $100$ for each stage.

\begin{table}[H]
\centering
\caption{Parameters of the target elasto-plastic curves for simultaneous design.}
\label{tab:plast_params}
\begin{tabular}{@{}cccccc@{}}
\toprule
Group & $E$ & $\varepsilon_y$ & $\sigma_{\infty}$ & $A_i$ & $\kappa_i$ \\
\midrule
1 & $4.8\times10^4$ & $2.7\times10^{-3}$ & 220 & [\ \ 4, 1, 85.4] & [1000, 88, 88] \\
2 & $6.8\times10^4$ & $2.8\times10^{-3}$ & 250 & [12, 1, 46.6] & [1000, 87, 87] \\
3 & $8.2\times10^4$ & $2.8\times10^{-3}$ & 280 & [18, 1, 31.4] & [1000, 63, 63] \\
\bottomrule
\end{tabular}
\end{table}

\begin{figure}[H]
    \centering
    {\includegraphics[width=1.0\textwidth]{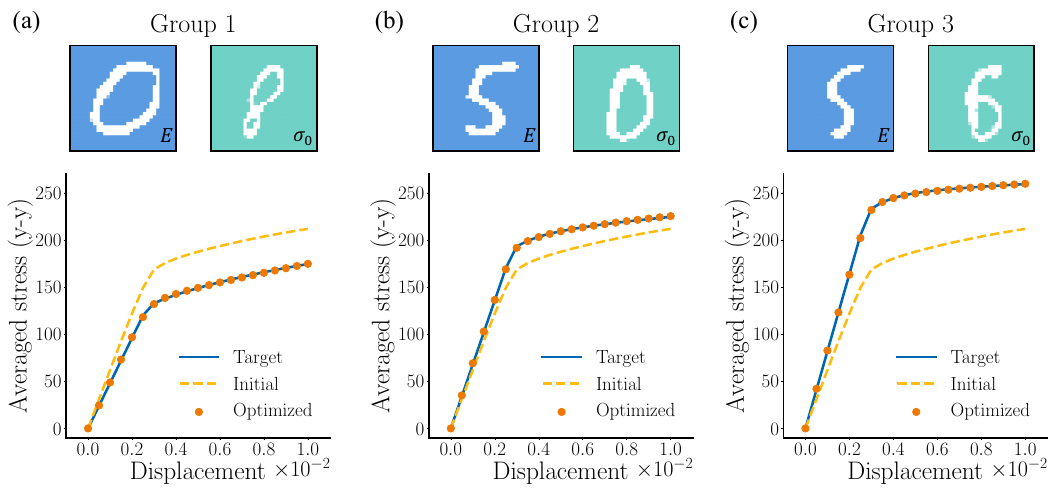}}
    \caption{Simultaneous design results of Young's modulus and yield stress fields to achieve different target elasto-plastic behaviors with the same initial designs: (a) Group 1; (b) Group 2; (c) Group 3.}
    \label{fig:plast_E_sig0}
\end{figure}

The simultaneous design results of Young's modulus and yield stress fields for the three groups are presented in Fig.~\ref{fig:plast_E_sig0}. The optimized microstructures exhibit distinct geometric features to achieve the different target properties. In Group $1$, the optimized Young's modulus field resembles the digit 0 with wider soft material regions to achieve a lower stiffness, while the yield stress field resembles the tiny digit 8 to achieve the continued increase in the post-yield stress. In Group $2$, the target curve is relatively close to the initial curve, resulting in optimized microstructures similar to the initial designs. In Group $3$, the optimized Young's modulus field retains the overall shape of the initial digit 5 but with a reduced scale to increase stiffness, and the yield stress field resembles digit 6 with a similar size to realize the slow increase in the post-yield stress. All the optimized microstructures achieve binarization metrics around $99\%$. All the stress-displacement curves of the optimized microstructures closely match the target curves, demonstrating the effectiveness of the proposed method in simultaneously designing multiple material property fields to achieve tailored elasto-plastic behaviors.

\subsection{Discussion}
\label{discussion}
The results presented in this section demonstrate the effectiveness of the proposed inverse design framework in generating microstructures with tailored mechanical properties while preserving stylistic features from the training data. Compared to existing inverse design methods based on conditional generative models, the proposed approach offers several advantages:

\begin{enumerate}
  \item Flexibility in design paradigm: 
  Existing conditional design methods aim to simultaneously generate microstructures with desired mechanical properties and style features by conditioning the reverse sampling on target properties. This paradigm typically involves preparing extensive datasets of microstructure-property pairs and designing sophisticated conditional embedding mechanisms. Moreover, any changes in the governing equations or boundary conditions typically require dataset regeneration and retraining of the conditional generative models, which can be time-consuming and computationally expensive. In contrast, the proposed method decouples the realization of tailored mechanical performance and stylistic features by reinterpreting the noise input as an optimizable design variable. This new paradigm allows for the diffusion model to focus solely on capturing the geometric features of the training data, while the optimization process effectively steers the design towards the desired mechanical properties. This flexibility enables the direct use of pre-trained unconditional diffusion models for diverse stylistic applications.
  
  \item Generality in target specification: 
  Owing to the optimization-based design process, the proposed framework can easily accommodate various forms of target performance specifications, including specific property values and entire response curves. This characteristic allows for greater versatility in addressing a wide range of design problems without the need for designing specialized conditional embedding mechanisms for different targets.
  
  \item Efficiency in implementation: 
  The unified differentiation pipeline via VJP concatenations enables efficient gradient evaluations through backpropagation. The customized VJP rule of the mechanical solver is problem-independent and can be easily extended to other mechanical systems. This capability significantly reduces the manual effort required for derivations and implementations, thereby accelerating the design process.

\end{enumerate}

Despite these advantages, current research also has some limitations that need investigation in future work:

\begin{enumerate}
  \item Degraded noise input: As observed in the hyperelastic design example, the optimized noise input tends to deviate from the isotropic Gaussian distribution when targeting extreme mechanical properties. 
  The degraded noise input can lead to generated microstructures deviating from the learned data distribution, resulting in a loss of stylistic features and a reduction in design diversity.
  Although these issues can be mitigated by selecting different combinations of base materials, future work could explore strategies to better regularize the noise input during optimization and enhance the generative capabilities of the diffusion model to capture a wider range of design variations.

  \item Soft geometric constraints: The generative model-based design methods can inherently capture the stylistic features of the training data but may not fully satisfy strict geometric constraints required in practical applications, as illustrated in the homogenization design example. Although the training dataset can satisfy these constraints, the generated samples may still violate them due to the nature of the generation process as well as the optimization procedure in our approach. Apart from the adopted multi-stage projection scheme to promote binarization and the discussed symmetry operations to enforce symmetric designs, future research could investigate the incorporation of additional constraints into the optimization formulation, such as the minimum feature size~\cite{lazarov2016length} and connectivity requirements~\cite{cool2025practical} that are well developed in the field of topology optimization.

\end{enumerate}

\section{Conclusion}
\label{sec:conclusion}
This work presents a new inverse design framework that integrates unconditional denoising diffusion models with advanced differentiable programming techniques to achieve the microstructure design with tailored mechanical properties and stylistic features. 
The reinterpretation of the noise input as a design variable transforms the traditional generative process into the solution of a constrained optimization problem, which enables the effective control of the reverse denoising process and eliminates the requirement for conditional embeddings. 
The developed unified differentiation pipeline via vector-Jacobian product concatenations enables efficient and automated sensitivity analysis through the backpropagation. 
The effectiveness of the proposed method is demonstrated through various microstructure design problems, including tailored homogenized, hyperelastic, and elasto-plastic properties. 
Numerical results show that the proposed approach can generate high-quality microstructures that closely match the desired mechanical properties while preserving stylistic features from the training data. 
This work offers a promising direction for integrating generative models with physics-based simulations for advanced material and structural design.

\section{Acknowledgement}

The authors acknowledge the Research Grant Council of Hong Kong for support through the ECS project (Ref No. 26205024).

\section{Data availability}
Our data and code are available at \href{https://github.com/CMSL-HKUST/genopt}{https://github.com/CMSL-HKUST/genopt}.


% \bibliographystyle{elsarticle-num}
% \bibliographystyle{elsarticle-harv}
% \bibliographystyle{elsarticle-num-names}
% \bibliographystyle{model1a-num-names}
% \bibliographystyle{model1b-num-names}
% \bibliographystyle{model1c-num-names}
% \bibliographystyle{model2-names}
% \bibliographystyle{model3a-num-names}
% \bibliographystyle{model3-num-names}
% \bibliographystyle{model4-names}
% \bibliographystyle{model5-names}
% \bibliographystyle{model6-num-names}
% \bibliographystyle{plainnat}

% \bibliography{refs}

\bibliographystyle{unsrt}
\bibliography{refs}

\end{document}